\documentclass[a4paper,11pt]{article}
\pdfoutput=1 

\usepackage{jheppub}

\usepackage{amsmath,amssymb,calc, amsthm,bbm, epsfig,psfrag, mathtools}
\usepackage{latexsym,bm,amsfonts}
\usepackage{graphicx, enumerate}
\usepackage{breqn}
\usepackage[numbers,sort&compress]{natbib}
\usepackage[dvipsnames]{xcolor}
\usepackage{tensor}
\usepackage{mathrsfs}

\allowdisplaybreaks

\newcommand{\Comment}[1]{{}}
\definecolor{darkblue}{rgb}{0.15,0.35,0.55}
\definecolor{reddish}{rgb}{0.65, 0.2, 0.2}
\usepackage[linktocpage=true]{hyperref}
\hypersetup{
colorlinks=true,
citecolor=darkblue,
linkcolor=reddish,
urlcolor=darkblue,
pdfauthor={},
pdftitle={},
pdfsubject={}
}

\makeatletter
\renewcommand\section{\@startsection {section}{1}{\z@}%
                                   {-3.5ex \@plus -1ex \@minus -.2ex}
                                   {2.3ex \@plus.2ex}%
                                   {\normalfont\large\bfseries}}
\renewcommand\subsection{\@startsection{subsection}{2}{\z@}%
                                     {-3.25ex\@plus -1ex \@minus -.2ex}%
                                     {1.5ex \@plus .2ex}%
                                     {\normalfont\bfseries}}
\makeatother

\let\non\nonumber

\newcommand{\bea}{\begin{eqnarray}}
\newcommand{\eea}{\end{eqnarray}}

\newcommand{\be}{\begin{equation}}
\newcommand{\ee}{\end{equation}}

\newcommand{\bma}{\begin{pmatrix}}
\newcommand{\ema}{\end{pmatrix}}

\newcommand{\C}[1]{$(\ref{#1})$}

\newcommand{\bsubeq}{\begin{subequations}}
\newcommand{\esubeq}{\end{subequations}}
\newcommand{\bsubea}{\begin{subequations}\bea}
\newcommand{\esubea}{\eea\end{subequations}}

\newfont{\goth}{ygoth.tfm scaled 1200}                   

 \numberwithin{equation}{section}

\def\1{{(1)}}
\def\2{{(2)}}
\def\3{{(3)}}

\def\s{\sigma}

\newcommand\Tb{\bar{T}}

\newcommand\calT{\mathcal{T}}
\newcommand\calTb{\bar{\mathcal{T}}}

\newcommand\Db{\bar{D}}

\def\a{\alpha}
\def\b{\beta}

\def\d{\delta}

\def\g{\gamma}

\def\l{\lambda}

\def\q{\theta}

\def\s{\sigma}

\def\z{\zeta}

\def\L{\Lambda}
\def\O{\Omega}

\def\S{\Sigma}

\newcommand{\ad}{{\dot\alpha}}

\newcommand{\pa}{\partial}
\newcommand{\qb}{{\bar{\theta}}}
\newcommand{\hf}{\frac12}

\newcommand {\cS}{{\cal S}}
\newcommand {\cN}{{\cal N}}

\newcommand {\cV}{{\cal V}}

\newcommand{\cO}{{\mathcal O}}
\newcommand{\cA}{{\mathcal A}}

\newcommand{\cQ}{{\mathcal Q}}
\newcommand{\cJ}{{\mathcal J}}
\newcommand{\cL}{{\mathcal L}}

\renewcommand{\be}{\begin{equation}}
\renewcommand{\ee}{\end{equation}}
\newcommand{\bpm}{\begin{pmatrix}}
\newcommand{\epm}{\end{pmatrix}}

\newcommand{\EV}[1]{\langle #1 \rangle}
\newcommand{\beqn}{\begin{eqnarray}}
\newcommand{\eeqn}{\end{eqnarray}}

\usepackage{slashed}
\newcommand{\cD}{\mathcal D}

\newcommand{\cR}{\mathcal R}

\newcommand{\sfD}{{\sf D}}
\newcommand{\sfF}{{\sf F}}

\newcommand{\ta}{\theta}

\newcommand{\tab}{\bar\theta}

\newcommand{\cX}{\mathcal X}

\newcommand{\p}{\partial}

\title{\boldmath  Non-Linear Supersymmetry and $T\bar T$-like Flows}

\author[a]{Christian Ferko,}
\author[b]{Hongliang Jiang,}
\author[a]{Savdeep Sethi,}
 
\author[b,c]{and \\ Gabriele Tartaglino-Mazzucchelli }

\affiliation[a]{Enrico Fermi Institute \& Kadanoff Center for Theoretical Physics, \\ 
University of Chicago, Chicago, IL 60637, USA}

\affiliation[b]{Albert Einstein Center for Fundamental Physics, Institute for Theoretical Physics,\\
University of Bern, Sidlerstrasse 5, CH-3012 Bern, Switzerland}

\affiliation[c]{School of Mathematics and Physics, University of Queensland\\
St Lucia, Brisbane, Queensland 4072, Australia}

\emailAdd{cferko@uchicago.edu}
\emailAdd{jiang@itp.unibe.ch}
\emailAdd{sethi@uchicago.edu}
\emailAdd{g.tartaglino-mazzucchelli@uq.edu.au}

\preprint{EFI-19-10}

\abstract{%
The $T\bar{T}$ deformation of a supersymmetric two-dimensional theory preserves the original supersymmetry. Moreover, in several interesting cases the deformed theory possesses additional non-linearly realized supersymmetries. We show this  
for certain $\cN =(2,2)$ models in two dimensions, where we observe an intriguing similarity with known $\cN=1$ models
in four dimensions. 
This suggests that higher-dimensional models with non-linearly realized supersymmetries might also be obtained from $T\bar{T}$-like flow equations. We show that in four dimensions this is indeed the case for \texorpdfstring{$\cN=1$}{N=1} Born-Infeld theory, as well as for the  Goldstino action for spontaneously broken \texorpdfstring{$\cN=1$}{N=1}  supersymmetry. 

}

\begin{document}

\maketitle
\flushbottom


\section{Introduction} 
\label{sec:intro}

There has been considerable recent excitement about quantum field theories in two dimensions deformed by the irrelevant operator $T\bar{T}$~\cite{Zamolodchikov:2004ce, Smirnov:2016lqw}. Part of the reason for excitement is that the deformed theory appears to be a new structure, which is neither a local quantum field theory nor a full-fledged string theory. There are many basic issues yet to be resolved, like how to define observables in the theory. What is understood, however, is the finite volume spectrum \cite{Smirnov:2016lqw,Cavaglia:2016oda}\ and the structure of the S-matrix \cite{Dubovsky:2013ira, Dubovsky:2017cnj}. For a recent overview, see the review \cite{Jiang:2019hxb}.

Another reason for excitement is apparent at the classical level. 
The $T\bar{T}$ deformation of a two-dimensional Lagrangian
leads to a classical flow equation for the deformed Lagrangian $\cL_\l(x)$ of the form
\begin{equation}
    \frac{d}{d\lambda}\mathcal{L}_{\lambda}
    =
    -\frac{1}{8}T \Tb
    \propto \det\big(T_{\mu\nu}[\mathcal{L}_{\lambda}]\big)
    \,,
    \label{sc1_flow}
\end{equation}
where $T_{\mu\nu}[\mathcal{L}_{\lambda}]$ is the stress-energy tensor for the deformed theory at value $\l$ of the flow 
parameter. When the undeformed theory is a free scalar theory, 
\be\label{freeaction}
S = {1\over 2}\int d^2x \, \partial_\mu \phi \partial^\mu \phi~, 
\ee
the deformed theory is the gauge-fixed Nambu-Goto string with string tension determined by the deformation parameter $\lambda$~\cite{Cavaglia:2016oda,Bonelli:2018kik}: 
\begin{align}
S &= \int d^2x \left( -{1\over 2 \lambda} + {1\over 2 \lambda}\sqrt{1+2\lambda \p_{\sigma}\phi \p^{\sigma}\phi}\right)~. 
\end{align}
This is a beautiful connection between $T\bar{T}$ deformations and a field theory which classically possesses a non-linearly realized $D=3$ Lorentz symmetry; for other connections between $T\bar{T}$ and classical string theory, see, for example,~\cite{Dubovsky:2012wk,Caselle:2013dra,Chen:2018keo, Dei:2018mfl, Baggio:2018gct, Frolov:2019nrr, Sfondrini:2019smd,Baggio:2018rpv}.

The magic of $T\bar{T}$ in two dimensions is that this composite operator is well-defined at the quantum level. This property does not currently extend to higher-dimensional candidates without some additional ingredients. One such potential ingredient is supersymmetry. Deforming a supersymmetric $D=2$ theory with $T\bar{T}$ preserves the original supersymmetry of the theory.
The supercurrent-squared operators that make the original supersymmetry manifest
have been explicitly constructed for various theories in   
\cite{Baggio:2018rpv, Chang:2018dge, Jiang:2019hux, Chang:2019kiu}. The usual $T\bar{T}$ operator is found as a supersymmetric descendant of supercurrent-squared 
up to equations of motion and total derivatives.

Some of the simplest examples studied so far are $T\bar T$ deformations
of supersymmetric free theories.
A remarkable feature of the deformed models is that the resulting interacting 
higher-derivative actions possess a set of hidden non-linear supersymmetries, in addition to their linearly realized ones. The deformed actions with  
$\mathcal N=(0,1),\,(1,1)$ and $(0,2)$  supersymmetry 
\cite{Baggio:2018rpv, Chang:2018dge, Jiang:2019hux}
coincide 
with gauge-fixed supersymmetric Nambu-Goto models, which exhibit 
various partial supersymmetry breaking patterns~\cite{Ivanov:2000nk}.

This connection between $T\bar{T}$ and structures which are central in string theory leads to a natural question: are more general classes of theories with non-linear symmetries related to flow equations for some analogue of $T\bar{T}$? One recent set of examples are the $\cN=(2,2)$ supersymmetric $T\bar T$-deformed actions of \cite{Chang:2019kiu}. Do they also admit non-linear supersymmetries? The answer is yes! 
Following  the ideas of \cite{Rocek:1997hi}, in this work we explicitly construct two models describing the partial supersymmetry 
breaking  pattern $\cN=(4,4)\rightarrow \cN=(2,2)$ in $D=2$. 
These models have manifest    $\cN=(2,2)$ supersymmetry from the superspace structure used in their construction, 
but they also admit another hidden non-linear $\cN=(2,2)$ supersymmetry. 
It turns out the resulting actions are exactly the same as the $\cN=(2,2)$  chiral and twisted chiral $T\bar T$-deformed actions of \cite{Chang:2019kiu}. 
The intriguing relation between non-linear supersymmetry and $T\bar T$ therefore persists for models with manifest $\cN=(2,2)$ supersymmetry. Interestingly, even the $D=2$  Volkov-Akulov action, describing the dynamics of the 
Goldstinos which arise from the spontaneous breaking of $\cN=(2,2)$ supersymmetry, satisfies 
 a $T\bar T$ flow equation~\cite{Cribiori:2019xzp}.

This collection of examples motivates us to see whether any higher-dimensional theories with non-linear supersymmetries might also satisfy $T\bar{T}$-like flow equations. It has been known for more than two decades that the Bagger-Galperin action for the $D=4$ $\cN=1$ Born-Infeld theory 
describes $\mathcal N=2\rightarrow \mathcal N=1$ partial supersymmetry breaking \cite{Bagger:1996wp}. Does 
the Bagger-Galperin action arise from a $T\bar T$-like deformation of $\cN=1$ Maxwell theory?
That the linear order deformation is given by a supercurrent-squared operator was noted long ago in~\cite{cecotti:1987}. Much more recently, bosonic Born-Infeld theory was shown to satisfy a $T^2$ flow equation, where $T^2$ is an operator quadratic in the stress-energy tensor~\cite{Conti:2018jho}. 
In this work, we explicitly show that the Bagger-Galperin action indeed satisfies a supercurrent-squared flow 
 equation, generalizing the observation of~\cite{cecotti:1987}  to all orders in the deformation parameter. The supercurrent-squared deformation operator is 
 constructed from supercurrent multiplets, but its top component contains other currents besides the stress-energy 
 tensor. This is different from $D=2$ where the top component of the supercurrent-squared operator is exactly the 
 standard $T\bar T$ operator on-shell.

This paper is organized as follows: in section~\ref{2D}, we show that $D=2$ $\cN=(2,2)$ deformed models of either free chiral or twisted chiral multiplets possess additional non-linearly realized $\cN=(2,2)$ supersymmetries. 
In section~\ref{Tsquaredcomments}, we describe a particular four-dimensional analogue of $T\bar{T}$ motivated by~\cite{Conti:2018jho}, and  generalize it to a supercurrent-squared operator for theories with $\cN=1$ supersymmetry. Section~\ref{bosonicBI} reviews the argument that relates bosonic Born-Infeld theory to the solution of a $T^2$ flow equation~\cite{Conti:2018jho}. In section~\ref{BI-flows}, we show that  $\cN=1$ Born-Infeld theory satisfies a supercurrent-squared flow equation 
to all orders in the deformation parameter.    
In section~\ref{Goldstino-Flow} we show that a particular form of the $D=4$ Goldstino action also satisfies a supercurrent-squared flow,  generalizing the $D=2$ result of \cite{Cribiori:2019xzp}. 
We end with concluding thoughts in section~\ref{concludingthoughts}. Appendix \ref{appendix:EoMBI} contains a useful result for the analysis of section \ref{BI-flows}.


\section{\texorpdfstring{$D=2 \;\, \cN=(2,2)$}{2D N=(2,2)} Flows and Non-Linear \texorpdfstring{$\cN=(2,2)$}{N=(2,2)} Supersymmetry} 
\label{2D}

 The $\cN=(2,2)$ supersymmetric extension of $T\bar{T}$ was recently studied in \cite{Chang:2019kiu}, where the existence of extra non-linearly realized supersymmetries for some solutions of the $T\bar{T}$ flow equation was briefly discussed.
In this section, we are going to explore in detail how these non-linear supersymmetries arise for the simplest $\cN=(2,2)$ $T\bar{T}$ flows. The undeformed models are supersymmetrized theories of free scalars, while the deformed models are $\cN=(2,2)$ supersymmetric extensions of the $D=4$ gauge-fixed Nambu-Goto string  studied in \cite{Chang:2019kiu}.
Before entering into the details of how the non-linear supersymmetry arises, let us review some of the results of \cite{Chang:2019kiu}
that are relevant for the analysis in this section.

\subsection{\texorpdfstring{$T \Tb$}{T Tbar} deformations with \texorpdfstring{$\cN=(2,2)$}{N=(2,2)} supersymmetry}

The composite operator
\bea
T \Tb(x)=T_{++++}(x)\,T_{----}(x)-\big[\Theta(x)\big]^2
~,
\label{component-TTbar}
\eea
written here in light-cone coordinates, possesses several remarkable features.
Although it is an irrelevant operator, it is quantum mechanically well-defined and preserves many of 
the symmetries of the undeformed theory \cite{Zamolodchikov:2004ce,Smirnov:2016lqw,Cavaglia:2016oda}.

In particular, $T\bar{T}$ deformations
preserve supersymmetry along the flow \cite{Baggio:2018rpv,Chang:2018dge,Jiang:2019hux,Chang:2019kiu, Coleman:2019dvf}. More specifically, the $T \Tb(x)$ operator of a supersymmetric theory is related to a supersymmetric 
descendant operator $\calT \calTb(x)$, 
\begin{align}
\calT\calTb(x)
=
T \Tb(x)
+{\rm EOM}
+\pa_{++}(\cdots)
+\pa_{--}(\cdots)
~.
\label{calTTb=TTb}
\end{align}
The previous equation states the equivalence of $T \Tb(x)$ and $\calT\calTb(x)$
 up to total derivatives and terms that vanish on-shell, which we have indicated with ``${\rm EOM}$''.
When $\cN=(2,2)$ supersymmetry is linearly realized and preserved along the flow, 
which is the case of interest for this analysis, 
$\calT \calTb(x)$ is expressed as a $D$-term, or full superspace integral, 
of a supercurrent-squared primary operator \cite{Chang:2019kiu}:
\be
\calT\calTb(x)
  = \int d^4 \theta \, 
  \mathcal{O}^{\rm FZ}(x,\q)
  ~,\quad
  \cO^{\rm FZ}(x,\q)
  :=
- \mathcal{J}_{++}(x,\q) \mathcal{J}_{--}(x,\q) + 2 \mathcal{V}(x,\q) \bar{\mathcal{V}}(x,\q)
~.
\ee
Here $\cJ_{\pm\pm}(x,\q)$, $\mathcal{V}(x,\q)$ and its complex conjugate $\bar{\mathcal{V}}(x,\q)$
are the local operators describing the Ferrara-Zumino (FZ) supercurrent multiplet 
for $D=2$ $\cN=(2,2)$ supersymmetry \cite{Ferrara:1974pz, Dumitrescu:2011iu}.\footnote{For simplicity, we have assumed 
that the $D=2$ $\cN=(2,2)$ theory possesses a well-defined FZ multiplet.
For a description of the more general case where one needs to use the $\cN=(2,2)$ $\cS$-multiplet of currents, discussed in
\cite{Dumitrescu:2011iu}, to define the the supercurrent-squared operator 
we refer to the original analysis of \cite{Chang:2019kiu}.}
These operators satisfy the following conservation equations
\be
    \Db_{\pm} \mathcal{J}_{\mp\mp} = \pm D_\mp \mathcal{V}  ~,\qquad
    \Db_\pm \cV=0~,
    \label{FZ-2}
\ee
together with their complex conjugates.
In superspace, assuming the supersymmetric Lagrangian $\cL_\l(x)$ along the flow is given by
\be
\mathcal{L}_\lambda(x) = \int d^4 \theta \, \mathcal{A}_\lambda(x,\q)
~,
\ee
with $\mathcal{A}_\lambda(x,\q)$ the full superspace Lagrangian,
the flow equation can be rewritten in a manifestly $\cN=(2,2)$ supersymmetric form:
\bea
    \frac{d}{d \lambda} \mathcal{A}_\lambda =- \frac{1}{8}\cO^{\rm FZ}= 
    \frac{1}{8} \left( \mathcal{J}_{++} \mathcal{J}_{--} 
    - 2 \mathcal{V} \bar{\mathcal{V}} \right)~ .
    \label{sc2_flow}
\eea

In \cite{Chang:2019kiu}  supersymmetric flows for various theories were studied. The simplest cases, on which we will focus in this section, 
 are $T\Tb$-deformed theories of free scalars, fermions and auxiliary fields.
In the case of $D=2$ $\cN=(2,2)$ supersymmetry, a scalar multiplet can have several different off-shell representations 
\cite{Gates:1984nk,Buscher:1987uw,Grisaru:1997pg,Lindstrom:2005zr}. The two cases we will consider here are chiral and twisted-chiral supermultiplets, which are the most commonly studied cases.

In $\cN=(2,2)$ superspace, parametrized by coordinates $\z^M=(x^{\pm\pm},\q^{\pm},\qb^\pm)$, let
the complex superfields $X(x,\q)$ and $Y(x,\q)$ satisfy chiral
and twisted-chiral constraints, respectively,
\bea
\bar  D_\pm X=0
~,\quad
\bar  D_+ Y= D_- Y=0
~.
\label{chiral_twisted-chiral_constraints}
\eea
Here the supercovariant derivatives and supercharges are\footnote{The reader should be aware that 
in this section we follow the notation of \cite{Ivanov:2004yv}, which is slightly different
from the notation used in \cite{Chang:2019kiu}.} 
\bsubea
 D_\pm &=&\frac{\p} {\p \theta^\pm}+i \bar \theta^\pm \p_{\pm\pm}
~, \quad
\bar  D_\pm =-\frac{\p} {\p\bar  \theta^\pm}  -  i  \theta^\pm \p_{\pm\pm}
~,
\\
 Q_\pm &=&i\frac{\p} {\p \theta^\pm}+\bar \theta^\pm \p_{\pm\pm}
~, \quad
\bar  Q_\pm =-i\frac{\p} {\p\bar  \theta^\pm}  -    \theta^\pm \p_{\pm\pm}
~,
\esubea
and they satisfy
\bsubea
 D_\pm^2&=&\bar  D_\pm^2=0
~, \qquad  
\{ D_\pm , \bar  D_\pm \}=-2i \p_{\pm\pm}
~, \qquad   
[D_\pm , \p_{\pm \pm} ]=[\bar D_\pm , \p_{\pm \pm} ]=0
~, \quad
\\
 Q_\pm^2&=&\bar  Q_\pm^2=0
~, \qquad  
\{ Q_\pm , \bar  Q_\pm \}=-2i \p_{\pm\pm}
~, \qquad   
[Q_\pm , \p_{\pm \pm} ]=[\bar Q_\pm , \p_{\pm \pm} ]=0
~. \quad
\esubea
There is one more caveat worth mentioning: in much of the $\cN=(2,2)$ literature, twisted-chiral multiplets, often denoted $\S$ in this context, naturally arise as field strengths for $\cN=(2,2)$ vector superfields $V$. The lowest component of such a superfield is a complex scalar, but the top component proportional to $\bar{\theta}^-\theta^+$ encodes the gauge-field strength along with a real auxiliary field. On the other hand, there are twisted chiral superfields denoted $Y$ whose bottom component is a complex scalar and whose top component is just a complex auxiliary field. It is to this latter case that we restrict. 
The free Lagrangians for these supermultiplets are given by
\bea
\mathcal{L}^{\text{c}}_{0} = \int d^4 \theta \, X\bar{X}
~,\qquad
\mathcal{L}^{\text{tc}}_{0} =
- \int d^4 \theta \, Y\bar{Y}
~.
\label{free-scaral-Lagrangians}
\eea

In \cite{Chang:2019kiu} it was shown that the following Lagrangian
\bsubea    \label{bgc_on_shell}
    \mathcal{L}^{\text{c}}_{\lambda} &=& \int d^4 \theta  \left( X \bar{X} 
    + \frac{\lambda  D_+ X \Db_+ \bar{X} D_-  X \Db_- \bar{X} }{1 - \frac{1}{2} \lambda  A 
    + \sqrt{1 - \lambda  A  + \frac{1}{4} \lambda^2  B^2 } } \right)~ ,
\eea
with
\bea
    A = \partial_{++}  X \partial_{--} \bar{X} + \partial_{++} \bar{X} \partial_{--}  X ~,
\quad
    B = \partial_{++}  X \partial_{--} \bar{X} - \partial_{++} \bar{X} \partial_{--}  X ~,
\esubea
is a solution of the flow equation \eqref{sc2_flow} on-shell, and hence describes the $T\Tb$-deformation
\eqref{sc1_flow} of the free chiral supermultiplet Lagrangian \eqref{free-scaral-Lagrangians}.

A simple way to generate
the $T\Tb$-deformation of the free twisted-chiral theory is to remember that a twisted-chiral multiplet can be obtained from a chiral one
by acting with a $\mathbb{Z}_2$ 
automorphism
on the Grassmann coordinates 
of $\cN=(2,2)$ superspace:
\bea
\q^+\leftrightarrow\q^+
~,\quad
\q^-\leftrightarrow-\,\bar\q^-
~.
\label{Z2auto}
\eea
This leaves the $D_+$ and $\bar{D}_+$ derivatives  invariant while it exchanges $D_-$ with $\bar{D}_-$. As a result, 
the chiral and twisted-chiral differential constraints \eqref{chiral_twisted-chiral_constraints} are mapped into each others under the automorphism \eqref{Z2auto}.\footnote{In the literature this ${\mathbb Z}_2$ automorphism
\eqref{Z2auto} is often called  the ``mirror-map'' or ``mirror-image" because it exchanges the vector and axial $U(1)$ R-symmetries.}

Under the $\mathbb{Z}_2$ 
automorphism \eqref{Z2auto}, the Lagrangian \eqref{bgc_on_shell} turns into the following twisted-chiral Lagrangian
\bsubea    \label{bgtc_on_shell}
    \mathcal{L}^{\text{tc}}_{\lambda} &=& -\int d^4 \theta  \left(Y \bar{Y}
    + \frac{\lambda  D_+ Y \Db_+ \bar{Y} \Db_- Y D_- \bar{Y} }{1 - \frac{1}{2} \lambda  A 
    + \sqrt{1 - \lambda  A  + \frac{1}{4} \lambda^2  B^2 } } \right)~ ,
   \eea
   where
   \bea
    A = \partial_{++} Y \partial_{--} \bar{Y} + \partial_{++} \bar{Y} \partial_{--} Y~,
\quad
    B = \partial_{++} Y \partial_{--} \bar{Y} - \partial_{++} \bar{Y} \partial_{--} Y ~. 
\esubea
Thanks to the map \eqref{Z2auto}, by construction the Lagrangian \eqref{bgtc_on_shell} is a $T\Tb$-deformation \eqref{sc1_flow}
and its superspace Lagrangian
$\cA^{\rm tc}_\l$,
$\cL^{\rm tc}_\l=\int d^4\q\,\cA^{\rm tc}_\l$, is an on-shell solution of the following flow equation
\bea
    \frac{d}{d \lambda} \mathcal{A}^{\text{tc}}_\lambda = 
    \frac{1}{8} \left( \mathcal{\cR}_{++} \mathcal{\cR}_{--} 
    - 2 \mathcal{B} \bar{\mathcal{B}} \right)~ .
    \label{sc2_flow_tc}
\eea
Here $\cR_{\pm\pm}(x,\q)$, $\mathcal{B}(x,\q)$ and its complex conjugate $\bar{\mathcal{B}}(x,\q)$
are the local operators describing the $\cR$-multiplet of currents
for $D=2$ $\cN=(2,2)$ supersymmetry 
that arise by applying \eqref{Z2auto} to the FZ multiplet of the chiral theory
\eqref{bgc_on_shell}~\cite{Chang:2019kiu}.
They satisfy the conservation equations, 
\be
    \Db_{+} \mathcal{R}_{--} 
    = i \Db_-{\cal B}
    ~,
\qquad
   D_{-} \mathcal{R}_{++} 
    = iD_+{\cal B}
    ~,\qquad
    \Db_+ {\cal B}=D_-{\cal B}=0~,
\label{conservation-R}
\ee
together with their complex conjugates.
Like the case of the FZ-multiplet, 
the supercurrent-squared operator
\be
\calT\calTb(x)
  = \int d^4 \theta \, 
  \mathcal{O}^{\cR}(x,\q)
  ~,\quad
  \cO^{\cR}(x,\q)
  :=
- \mathcal{R}_{++}(x,\q) \mathcal{R}_{--}(x,\q) + 2 \mathcal{B}(x,\q) \bar{\mathcal{B}}(x,\q)
~,
\ee
satisfies \eqref{calTTb=TTb}; namely, $\calT\calTb(x)$
is equivalent to $T\bar{T}(x)$ up to total derivatives and ${\rm EOM}$  \cite{Chang:2019kiu}.

Note that the bosonic truncation of both \eqref{bgc_on_shell} and \eqref{bgtc_on_shell} give the Lagrangian
\begin{equation}
\label{eq:bosonTTbar}
 {\cal L}_{\lambda,\text{bos}}= 
 \frac{\sqrt{1+2 \lambda  a +\lambda^2 b^2}-1}{4\lambda}
 =
  \frac{  a}{4 } - \lambda \frac{ \pa_{++} \phi \pa_{--}\phi \pa_{++} \bar \phi \pa_{--} \bar\phi }
 {  1+\lambda a+\sqrt{1+2 \lambda  a +\lambda^2 b^2}  } 
 ~,
\end{equation}
where
\begin{equation}
  \label{eq:xydef}
 a=\pa_{++} \phi \pa_{--} \bar \phi +\pa_{++}  \bar \phi \pa_{--}  \phi
 ~, \qquad 
b= \pa_{++} \phi \pa_{--} \bar \phi -\pa_{++}  \bar \phi \pa_{--}  \phi ~,
\end{equation}
and $\phi$ is either $\phi= X|_{\q=0}$ or $\phi=Y|_{\q=0}$.
This is the Lagrangian for the gauge-fixed Nambu-Goto string in four dimensions
\cite{Cavaglia:2016oda}.

The aim of the remainder of this section is to show
that the Lagrangians \eqref{bgc_on_shell} and \eqref{bgtc_on_shell}
are structurally identical to the Bagger--Galperin action
for the $D=4$ $\cN=1$ supersymmetric Born-Infeld theory 
\cite{Bagger:1996wp}, which we will analyse in detail in section \ref{BI-flows}. 
Since the Bagger--Galperin action possesses a second non-linearly realized $D=4$ $\cN=1$ supersymmetry,
we will show that the theories described by \eqref{bgc_on_shell} and \eqref{bgtc_on_shell} also possess an extra set of non-linearly realized $\cN=(2,2)$ supersymmetries.

\subsection{The \texorpdfstring{$T \Tb$}{T Tbar}-deformed twisted-chiral model and  partial-breaking}

Let us start with the twisted-chiral Lagrangian \eqref{bgtc_on_shell}
which, as we will show, is the one more directly related to the $D=4$ Bagger-Galperin action.
In complete analogy to the $D=4$ case, we are going to show that \eqref{bgtc_on_shell} is a model for a Nambu-Goldstone multiplet of $D=2$ 
$\cN=(4,4)\to \cN=(2,2)$ partial supersymmetry breaking.
The analysis is similar in spirit to the $D=4$ construction of the Bagger-Galperin action using $D=4$ $\cN=2$ superspace proposed by Ro\v{c}ek and Tseytlin  \cite{Rocek:1997hi}; see also
\cite{Kuzenko:2015rfx,Antoniadis:2017jsk,Antoniadis:2019gbd} for  more recent analysis.

To describe manifest $\cN=(4,4)$ supersymmetry we can use $\cN=(4,4)$ superspace which augments the $\cN=(2,2)$ superspace coordinates $\z^M=(x^{\pm\pm},\q^{\pm},  \bar \q^\pm)$  of the previous section with
the following additional complex Grassmann coordinates $(\eta^{\pm},  \bar \eta^\pm)$.  
The extra supercovariant derivatives and supercharges are given by
\bsubeq
\bea
 \cD_+ 
 &=&
 \frac{\p} {\p \eta^+ }+i \bar \eta^+ \p_{++ }
 ~, \quad
\bar \cD_+ =-\frac{\p} {\p \bar\eta^+ } - i  \eta^+ \p_{++ }
~,
\\
 \cQ_+ 
 &=&
 i \frac{\p} {\p \eta^+ }+  \bar \eta^+ \p_{++ }
 ~, \quad
\bar \cQ_+ =- i \frac{\p} {\p \bar\eta^+ } -    \eta^+ \p_{++ }
~,
\eea
 \esubeq
with similar expressions for $\cD_-$ and $\cQ_-$. They satisfy
\bsubea
 \cD_\pm^2&=&\bar  \cD_\pm^2=0
~, \qquad  
\{ \cD_\pm , \bar  \cD_\pm \}=-2i \p_{\pm\pm}
~, \qquad   
[\cD_\pm , \p_{\pm \pm} ]=[\bar \cD_\pm , \p_{\pm \pm} ]=0
~,~~~~~~
\\
 \cQ_\pm^2&=&\bar  \cQ_\pm^2=0
~, \qquad  
\{ \cQ_\pm , \bar  \cQ_\pm \}=-2i \p_{\pm\pm}
~, \qquad   
[\cQ_\pm , \p_{\pm \pm} ]=[\bar \cQ_\pm , \p_{\pm \pm} ]=0
~,~~~~~~
\esubea
while they (anti-)commute with all the usual $D_\pm$ and $Q_{\pm}$ operators.

Two-dimensional $\cN=(4,4)$ supersymmetry can also be usefully described in the language of $\cN=(2,2)$ superspace.
In this section, we will largely refer to \cite{Ivanov:2004yv} for such a description. 
In this approach from the full $(4,4)$ supersymmetry, one copy of $(2,2)$ is manifest while a second $(2,2)$ is hidden. 
For our goal of describing a model of partial supersymmetry breaking, 
we view the hidden $(2,2)$ supersymmetry as broken and non-linearly realized. 
We will derive such a description starting from $\cN=(4,4)$ superspace and describe the broken/hidden supersymmetry 
using the $\eta^{\pm}$ directions.

The hidden supersymmetry transformation of a generic $D=2$ $\cN=(4,4)$ superfield 
$U=U(x^{\pm\pm},\q^\pm,\qb^\pm,\eta^\pm,\bar\eta^\pm)$ under the hidden $(2,2)$ supersymmetry  is
 \be
 \delta U= i(\epsilon^+ \cQ_+ +\epsilon^- \cQ_- - \bar \epsilon^+  \bar \cQ_+-\bar \epsilon^- \bar \cQ_-  )U
 ~.
 \label{hidden22_1}
 \ee
The $(2,2)$ supersymmetry, generated by the $Q_\pm$ and $\bar{Q}_\pm$ operators,
will always be manifest and preserved,
so we will not bother to discuss it in detail. 
For convenience, we also introduce the chiral coordinate $y^{\pm\pm}=x^{\pm\pm} + i \eta^\pm \bar \eta^\pm $.
Using this coordinate, the spinor covariant derivatives and supercharges take the form
\bsubea
 \cD_\pm 
 &=&
 \frac{\p} {\p \eta^\pm } + 2 i  \bar\eta^\pm \frac{\p}{\p y^ {\pm\pm }}
 ~,  \quad
\bar \cD_\pm =-\frac{\p} {\p \bar\eta^\pm }
~,
\\
 \cQ_\pm &=&
 i \frac{\p} {\p \eta^\pm } ~, \quad
\bar \cQ_\pm =- i \frac{\p} {\p \bar\eta^\pm }  - 2  \eta^\pm\frac{\p}{\p y^ {\pm\pm }} 
~.
\esubea

After this technical introduction, let us turn to our main construction.
  Consider a $(4,4)$  superfield   which is   chiral under the hidden $(2,2)$ supersymmetry: 
 \be
\bar  \cD_\pm {  \bm{\mathcal X}} =0
~.
 \ee
We can  expand it in terms of   hidden fermionic coordinates, 
\be
  {\bm{\mathcal X}} =X +  \eta^+ X_++  \eta^- X_- +   \eta^+  \eta^-F
  ~,
  \label{calX}
\ee
 where $X=X(y^{\pm\pm},\q^\pm,\bar\q^\pm)$,
 $X_\pm=X_\pm(y^{\pm\pm},\q^\pm,\bar\q^\pm)$
 and $F=F(y^{\pm\pm},\q^\pm,\bar\q^\pm)$ 
 are   themselves $(2,2)$
 superfields. In the following discussion, we will keep
 the $\theta^\pm ,\,\bar \theta^\pm$ dependence implicit. 
 The hidden $(2,2)$ supersymmetry transformation rules 
 can then be straightforwardly computed using \eqref{hidden22_1} and \eqref{calX}. 
 They take the form 
 \bsubeq
 \label{susyYsfXY}
 \bea
 \delta X &=&  - \epsilon^+ X_+   - \epsilon^- X_-  
 ~,
 \label{susyYsfXY_a}
 \\
 \d X_\pm&=& \mp \epsilon^\mp F -2i \bar \epsilon ^\pm \p_{\pm \pm }X,
  \label{susyYsfXY_b}
 \\
  \delta F &=&  - 2i \bar \epsilon^-  \p_{--} X_+ + 2i \bar \epsilon^+ \p_{++} X_-
  ~.
   \label{susyYsfXY_c}
\eea
\esubeq
The $\bm{\mathcal X}$ superfield is still reducible under $\cN=(4,4)$ supersymmetry so we can put additional constraints on the $(2,2)$ superfields $X$, $X_\pm$ and $F$. 
Here we will consider $(4,4)$ twisted multiplets, and refer the reader to \cite{Gates:1984nk,Gates:1983py,Gates:1995aj,Gates:1998fr,Ivanov:1984ht,Ivanov:1984fe,Ivanov:1987mz} for a more detailed analysis.
For this discussion, we will follow the $\cN=(2,2)$ superspace description of \cite{Ivanov:2004yv}.
One type of twisted multiplet with $(4,4)$ supersymmetry can be defined by setting  
\be
X_+= \bar  D_+ \bar Y, \qquad X_-=  -\bar  D_-  Y
~,
\ee
where  $X$ and $Y$ are chiral and twisted-chiral, respectively, under the manifest $(2,2)$ supersymmetry: 
\be\label{XYchiral}
\bar  D_+ X=\bar  D_- X=\bar  D_+ Y =   D_- Y=0
~, \qquad
  D_+ \bar  X=   D_- \bar  X=   D_+ \bar  Y = \bar   D_-  \bar  Y=0
 ~.
\ee
The superfield \eqref{calX} becomes
\be
{{\bm{\mathcal X}} }=X+   \eta^+    \Db_+ \bar Y  - \eta^- \bar  D_- Y+    \eta^+ \eta^-F
~.
\ee
The supersymmetry transformation rules then become
 \bsubea\label{susyYsfXY-2}
 \delta X &=&  - \epsilon^+  \bar  D_+ \bar Y   + \epsilon^-  \bar  D_-  Y 
 ~,
 \label{susyYsfXY-2_X}
\\
  \delta F &=&  - 2i \bar \epsilon^-  \p_{--}  \bar  D_+ \bar Y - 2i \bar \epsilon^+ \p_{++}   \bar  D_-  Y
  ~,
 \esubea
 while $\d X_\pm$ remains the same as \eqref{susyYsfXY_b}.
By using the conjugation property for two fermions,  
$\overline {\chi\xi} =\bar\xi\bar \chi =-\bar\chi \bar \xi $, 
and the conjugation property $\overline{ D_+ A} =\bar  D_+ \bar A $ 
for a bosonic superfield $A$, it follows that
 \be
  \delta \bar X =    \bar \epsilon^+   D_+ Y   -  \bar \epsilon^-    D_-  \bar  Y 
  ~.
 \ee
  One can check that
\be
 \delta  \bar  D^2 \bar X =\bar  D^2 \delta  \bar X
 =\bar  D^2 \Big(    \bar \epsilon^+   D_+ Y   -  \bar \epsilon^-    D_-  \bar  Y    \Big) 
 =2i \bar\epsilon^+ \p_{++} \bar  D_- Y +2i \bar \epsilon^- \p_{--}\bar D_+ Y
 ~,
\ee
where $\bar  D^2 =\bar  D_+ \bar  D_-$. Note that in the first equality we made use of the fact that the manifest and hidden 
$(2,2)$ supersymmetries are independent.   The supersymmetry transformation rule for $-\bar  D^2 X$ is then exactly that of the auxiliary field $F$. Thus we can  consistently set
\bea
F= -\bar  D^2\bar X
~,
\eea
which is the last constraint necessary to describe a version of the $(4,4)$ twisted multiplet
in terms of a chiral and twisted-chiral $\cN=(2,2)$ superfields.
The resulting  $(4,4)$ superfield ${{\bm{\mathcal X}} }$,
expanded in terms of the hidden $(2,2)$ fermionic coordinates, takes the form
\be
{{\bm{\mathcal X}} }=X+   \eta^+    \Db_+ \bar Y  - \eta^- \bar  D_- Y  -    \eta^+ \eta^-\bar  D^2\bar X
~,
\label{calX1}
\ee
which closely resembles the expansion of a $D=4$ $\cN=2$ vector multiplet when one identifies 
the analogue of the $D=4$ $\cN=1$ chiral vector multiplet field strength $W_\a$ with the $(2,2)$ chiral superfields
$\Db_+\bar{Y}$
and
$\Db_- Y$.
Note in particular that ${{\bm{\mathcal X}} }$ turns to be $(4,4)$ chiral:
 \be
\bar  D_\pm {  \bm{\mathcal X}} =0
~,\quad
\bar  \cD_\pm {  \bm{\mathcal X}} =0
~.
 \ee
To summarize: the entire $(4,4)$ off-shell twisted multiplet is described in terms of one chiral and one twisted-chiral $(2,2)$ superfield, which 
possess the following hidden $(2,2)$ supersymmetry transformations:
 \bsubea\label{susyYsfXY-3}
 \delta X &=&  - \epsilon^+  \bar  D_+ \bar Y   + \epsilon^-  \bar  D_-  Y 
 \label{susyYsfXY-3-X}
 ~,\\
 \d Y&=& \bar \epsilon^- D_- X+\epsilon^+ \bar D_+ \bar X
  ~.
 \esubea

Let us now introduce the action for a free $\cN=(4,4)$ twisted multiplet.
Taking the square of ${{\bm{\mathcal X}} }$ in \eqref{calX1} we obtain
\be
{ {\bm{\mathcal X}} } ^2= \eta^+  \eta^-  \Big( -2X   \bar  D^2 \bar X+2  \bar  D_+ \bar Y \bar  D_- Y \Big) +\ldots
~,
\ee
where the ellipses denote terms that are not important for our analysis. 
Since ${\bm{\mathcal X}}  $ and therefore ${\bm{\mathcal X}} ^2$ are chiral superfields, we can consider the chiral integral in the hidden 
direction
\be
\int d \eta^+ d  \eta^-{{\bm{\mathcal X}} } ^2 =2X   \bar  D^2 \bar X-2  \bar  D_+ \bar Y \cdot  \bar  D_- Y
~.
\ee
 Note also that,  since $X$ and $Y$ are chiral and twisted-chiral  under the manifest supersymmetry \eqref{XYchiral}, it follows that 
\bea
\int d^2x\,d\theta^+ d\theta^- d\bar \theta^+ d\bar \theta^- (X\bar X-Y \bar Y)
&=&\int d^2x\,d\theta^+ d\theta^-  \bar  D _+  \bar  D_- (X\bar X-Y \bar Y)
~,
\non\\
&=&\int d^2x\,d\theta^+ d\theta^-   \Big(X \bar  D _+  \bar  D_-\bar X-  \bar  D_+ \bar Y \cdot  \bar  D_- Y 
 \Big)
 ~,~~~~~~
\eea
which can also be rewritten as 
\be
\int d^2x\,d\theta^+ d\theta^- d\bar \theta^+ d\bar \theta^- (X\bar X-Y \bar Y)
=
\int d^2x\,d\bar\theta^+ d\bar\theta^-   \Big(\bar X    D _+     D_-  X-    D_+  Y \cdot     D_- \bar Y  \Big) 
~.
\ee
The sum of the two equations above  yields
\be\label{action44SUSY}
4 \int d^2x\,d\theta^+ d\theta^- d\bar \theta^+ d\bar \theta^- (X\bar X-Y \bar Y) 
= \int d^2x\,d\theta^+ d\theta^-    d \eta^+ d  \eta^-{ {\bm{\mathcal X}} } ^2 +{}c.c.~.
\ee
The left-hand side has an enhanced $\cN=(4,4)$ supersymmetry as discussed in  \cite{Ivanov:2004yv}. This becomes manifest from our $(4,4)$ superspace construction on the right-hand side.

To describe $\cN=(4,4)\to\cN=(2,2)$ supersymmetry breaking
we can appropriately deform the $(4,4)$ twisted multiplet. 
Analogous to the case of a $D=4$ $\cN=2$ vector multiplet deformed by a magnetic Fayet-Iliopoulos term \cite{Antoniadis:1995vb}
(see also \cite{Antoniadis:2019gbd,Rocek:1997hi,Ivanov:1998jq,Kuzenko:2015rfx,Antoniadis:2017jsk}), 
we add a deformation parameter to the auxiliary field $F$ of ${{\bm{\mathcal X}} }$,
which is deformed to
\be
{{\bm{\mathcal X}} }_{\text{def}}=X+   \eta^+    D_+ \bar Y  - \eta^- \bar  D_- Y  -    \eta^+ \eta^- \Big( \bar  D^2\bar X+\kappa\Big)
~.
\ee
Assuming that the auxiliary field $F$ gets a VEV, $\langle F\rangle=\kappa$ or equivalently $\langle \Db^2\bar X\rangle=0$,
then by looking at the supersymmetry transformations of $X_\pm$ for the deformed multiplet
\bea
\d X_\pm
&=&
  \pm \epsilon^\mp \Big( \bar  D^2\bar X+\kappa\Big)
  -2i \bar \epsilon ^\pm \p_{\pm \pm }X
~,
\label{dYdef}
\eea
we can see the $\cN=(4,4)\to\cN=(2,2)$ supersymmetry breaking pattern arises; specifically, the hidden $\cN=(2,2)$ is spontaneously broken and non-linearly realized.
For later use, it is important to stress that, though the hidden transformations of $\d X_\pm$ are modified by the non-linear term
proportional to $\kappa$,
the hidden transformation of $X$ remains the same as in the undeformed case given in eq.~\eqref{susyYsfXY-2_X}.

In analogy to the $D=4$ case of \cite{Rocek:1997hi,Kuzenko:2015rfx,Antoniadis:2017jsk}, 
to describe the Goldstone multiplet associated to partial supersymmetry breaking
we impose the following nilpotent constraint on the deformed $(4,4)$ twisted superfield:
\be
{ {\bm{\mathcal X}} }_{\text{def}} ^2=0=-2   \eta^+  \eta^-  \Big(    X (\kappa+   \bar  D^2 \bar X )-  \bar  D_+ \bar Y  \cdot \bar  D_- Y \Big) 
+\ldots
~.
\ee
This implies the constraint 
\be
 X  \Big(\kappa +   \bar  D^2  \bar X \Big)-  \bar  D_+ \bar Y \cdot  \bar  D_- Y=0
 ~,
 \label{BG44-1}
\ee
which requires
\be\label{Xeq}
X=\frac{ \bar  D_+ \bar Y \cdot  \bar  D_- Y}{\kappa+   \bar  D^2\bar  X}  =\frac{ W^2}{ {\kappa}+   \bar  D^2\bar  X}
~,
\ee
and its conjugate 
\be\label{Xbeq}
\bar{X}=
-\frac{  D_+ Y \cdot   D_- \bar{Y}}{\kappa+   D^2X}
= \frac{     \bar W^2 }{ {\kappa}+     D^2  X}
~.
\ee
Here $\bar  D^2 =\bar  D_+ \bar  D_-,   D^2 =-  D_+   D_-$ and we have  introduced the superfields:
\bsubea
W^2 
&=& 
   -X_+ X_-
   =   \bar  D_+ \bar Y \cdot  \bar  D_- Y 
   =\bar  D_+      \bar  D_-  (Y \bar Y)= \bar D^2 (Y\bar Y)~,
   \\
\bar W^2 &=&
   \bar{X}_+ \bar X_-
   =   -   D_+   Y \cdot   D_- \bar Y 
   = -  D_+   D_-  ( Y\bar Y) =   D^2 (Y\bar Y)
~.
\esubea
The constraint \eqref{BG44-1} is the $D=2$ analogue of the Bagger-Galperin constraint for a Maxwell-Goldstone multiplet for 
$D=4\,$ $\cN=2\to\cN=1$ supersymmetry breaking \cite{Bagger:1996wp}. 
 Combining \eqref{Xeq} and \eqref{Xbeq} gives 
\be\label{XwithY}
\kappa X=\bar D^2 (   Y\bar Y - X\bar X)=\bar D^2 \Big[ Y\bar Y - \frac{   \bar  D_+ \bar Y \cdot  \bar  D_- Y\cdot    D_- \bar Y \cdot    D_+  Y  }{
( {\kappa}+     D^2  X)( {\kappa}+   \bar  D^2  \bar X)}   \Big]
~,
\ee
which is consistent thanks to the $\kappa$ terms in the denominator.
Because of the four fermion coupling in the numerator of the last term, 
 no fermionic terms can appear in the denominator. So effectively we have the equation
\be
(\kappa + D^2 X)_{\text{eff}}  =  \Big(\kappa + D^2  \frac{ W^2}{ {\kappa}+   \bar  D^2\bar  X}\Big)_{\text{eff}}
=\kappa+\frac{ D^2 W^2}{\kappa+(\bar D^2 \bar X)_{\text{eff}}}
~,
\ee
and its conjugate 
\be
(\kappa +\bar D^2\bar X)_{\text{eff}} = \kappa+\frac{ \bar  D^2 \bar W^2}{\kappa+(   D^2 X)_{\text{eff}}}
~.
\ee
Solving them we get 
\bsubea
(   D^2  X)_{\text{eff}} &=&\frac{B-\kappa^2 +\sqrt{B^2 +2\kappa^2 A+\kappa^4}   }{2\kappa}
~, 
\\
( \bar D^2\bar X)_{\text{eff}} &=&\frac{-B-\kappa^2 +\sqrt{B^2 + 2\kappa^2 A+\kappa^4}   }{2\kappa}
~.
\esubea
Substituting these expressions into \eqref{XwithY} gives
\be
X=\frac{1}{\kappa} \bar D^2\Upsilon
~, \qquad 
\bar  X=\frac{1}{\kappa}   D^2 \Upsilon
~, \qquad 
 \Upsilon=\bar\Upsilon= Y\bar Y - \frac{2 W^2 \bar W^2}{A+\kappa^2 +\sqrt{B^2 + 2\kappa^2 A+\kappa^4}   }  
 ~,
\ee
where 
\bsubeq\label{2DAB}
\bea
A&=& D^2 W^2 +\bar D^2 \bar W^2=\{  D^2, \bar  D^2 \}(Y\bar Y)
=\partial_{++} Y \partial_{--} \bar{Y} + \partial_{++} \bar{Y} \partial_{--} Y
~ ,
\\
B&=&
 D^2 W^2 -\bar D^2 \bar W^2= [  D^2, \bar  D^2  ](Y\bar Y)
 =\partial_{++} Y \partial_{--} \bar{Y} - \partial_{++} \bar{Y} \partial_{--} Y
~.
\eea
\esubeq
The result is that the $\cN=(2,2)$ chiral part  $X$ of the $\cN=(4,4)$ twisted multiplet is expressed 
in terms of the $(2,2)$ twisted-chiral superfield $Y$. 
Thanks to the linearly realized construction in terms of $(4,4)$ superfields, it is straightforward
to obtain the non-linearly realized $\cN=(2,2)$ supersymmetry transformations for $Y$. In particular, 
it suffices to look at the transformations of
$D_+Y$ and $\Db_- Y$ that
can be obtained by substituting back the composite expression for $X=X[Y]$ into the 
transformations \eqref{dYdef}. By construction, these expressions ensure that $\d X$ transforms according to \eqref{susyYsfXY-2_X}.

Since $X$ is chiral under the manifest $(2,2)$ supersymmetry \eqref{XYchiral}, we can consider the chiral  integral
\bea
S_{\kappa^2}&=&
-\frac12 \kappa\int d^2 x \,d\theta^+ d\theta^- X+c.c
=-\frac12  \int d^2x\,d\theta^+ d\theta^-   \bar D^2\Upsilon +c.c.
\non\\
&=&-  \int d^2x\,d\theta^+ d\theta^- d \bar \theta^+ d\bar \theta^-   \Upsilon  
 ~.
\eea
A remarkable property of this action is that
it is invariant under the hidden non-linearly realized supersymmetry. 
Using \eqref{susyYsfXY-2_X}, we see that 
\bsubeq
\label{deriv-susy-1}
\bea
\delta S_{\kappa^2} &= &
-\frac12 \kappa\int d^2 x D_+ D_- \delta X\Big|_{\theta=\bar\theta=0}+c.c.~,
 \\
 &= &
 - \frac12 \kappa\int d^2 x  \Big( -2i  \epsilon^-  \p_{--} D_+  Y 
-2i  \epsilon^+ \p_{++}     D_- \bar Y\Big) \Big|_{\theta=\bar\theta=0}+c.c.=0
~,
\eea
\esubeq
where we used the fact that $Y$ is a twisted-chiral superfield \eqref{XYchiral}.  

Explicitly, the action reads 
\be
S_{\kappa^2}= -\int d^2x\,d\theta^+ d\theta^- d \bar \theta^+ d\bar \theta^-  
\Bigg( Y\bar Y - \frac{  2W^2 \bar W^2}{\kappa^2+A+ \sqrt{\kappa^4+2 \kappa^2 A+ B^2  }   } \Bigg)
~,
\ee
which precisely matches the model of eq.~\eqref{bgtc_on_shell} if we 
identify the coupling constants:
\be
\l =-\frac{2}{\kappa^2}
~.
\ee
This shows explicitly that the $T\Tb$-deformation of the free twisted-chiral action possesses a non-linearly realized 
$\cN=(2,2)$ hidden 
supersymmetry.

\subsection{The \texorpdfstring{$T \Tb$}{T Tbar}-deformed chiral model and  partial-breaking}

Let us now turn to the $T\Tb$ deformation of the free chiral model of eq.~\eqref{bgc_on_shell}. The construction follows the previous
subsection with the difference that we will start with a different formulation of the $(4,4)$ twisted multiplet
described in terms of $(2,2)$ superfields.
Consider again an $\cN=(4,4)$  superfield   which is   chiral under the hidden (2,2) supersymmetry: 
 \be
\bar  \cD_+ {  \bm{\mathcal Y}} =\bar \cD_-{  {\bm{\mathcal Y}} }=0
~.
 \ee
Its expansion in hidden superspace variables is  
\be
  {\bm{\mathcal Y}}=Y +   \eta^+ Y_++   \eta^- Y_- +   \eta^+  \eta^-G
  ~,
\ee
 where $Y=Y(y^{\pm\pm},\q^\pm,\bar{\q}^{\pm})$, $Y_\pm=Y_\pm(y^{\pm\pm},\q^\pm,\bar{\q}^{\pm})$  
 and $G=G(y^{\pm\pm},\q^\pm,\bar{\q}^{\pm})$ 
 are themselves superfields with manifest $(2,2)$ supersymmetry.
 The hidden $(2,2)$ supersymmetry transformation rules of the components are 
\bsubea\label{susyYsfXY_4}
 \delta Y &=&  - \epsilon^+ Y_+   - \epsilon^- Y_-  
 ~,\\
 \d Y_\pm&=&
  \mp \epsilon^\mp G -2i \bar \epsilon ^\pm \p_{\pm \pm }Y
 ~,
 \\
  \delta G &=&  - 2i \bar \epsilon^-  \p_{--} Y_+ + 2i \bar \epsilon^+ \p_{++} Y_-
  ~.
\esubea

This representation of $(4,4)$  off-shell supersymmetry is again reducible so we can impose constraints. As in the construction of the previous section, we impose
\be
Y_+= \bar  D_+ \bar X
~, \qquad Y_-=   D_-  X
~,
\ee
then
\be
  {\bm{\mathcal Y}} =Y +   \eta^+ \bar  D_+ \bar X  +   \eta^-   D_-  X  +   \eta^+  \eta^-G
  ~.
\ee
Here  $X$ and $Y$ are consistently chosen to be chiral and twisted-chiral under the manifest $(2,2)$ supersymmetry:
\be\label{XYchiral-2}
\bar  D_+ X=\bar  D_- X=\bar  D_+ Y =   D_- Y=0
~, \qquad
  D_+ \bar  X=   D_- \bar  X=   D_+ \bar  Y = \bar   D_-  \bar  Y=0
 ~. 
\ee
Then we have 
\be
\delta Y =-\epsilon^+ \bar  D_+ \bar X-\epsilon^-  D_- X
~,
\label{dY2}
\ee
as well as its conjugate 
\be
\delta \bar Y= \bar \epsilon^+  D_+ X +\bar\epsilon^- \bar  D_- \bar X
~.
\ee
Hence it follows that
\be
\delta (\bar D_+  D_- \bar Y) =\bar D_+  D_- \delta  \bar Y
=2i  \bar \epsilon^+ \p_{++}  D_- X - 2i \bar\epsilon^- \p_{--} \bar D_+ \bar X
~.
\ee
This should be compared with
\be
\delta G=2i  \bar \epsilon^+ \p_{++}  D_- X - 2i \bar\epsilon^- \p_{--} \bar D_+ \bar X
~,
\ee
showing that $\bar D_+  D_- \bar Y$ transforms exactly like the auxiliary field $G$. 
This enables us to further constrain the $(4,4)$ multiplet  by setting 
\be
G= \bar D_+  D_- \bar Y
~.
\ee
Imposing these conditions gives a $(4,4)$ twisted superfield 
\be
  {\bm{\mathcal Y}} =Y +   \eta^+ \bar  D_+ \bar X  +   \eta^-   D_-  X  +   \eta^+  \eta^-  \bar D_+  D_- \bar Y
  ~,
\ee
which by construction is twisted-chiral and chiral with respect to the manifest and hidden $(2,2)$ supersymmetries, respectively:
\be
\bar  D_+ {  \bm{\mathcal Y}} =
D_- {  \bm{\mathcal Y}} =0
~,\quad
\bar  \cD_\pm {  \bm{\mathcal Y}} =0
~.
\ee
Its free dynamical action can be easily constructed by considering  its square
\be
  {\bm{\mathcal Y}}^2 =2  \eta^+ \eta^- \Big(  Y  \bar D_+  D_- \bar Y - \bar  D_+ \bar X   \cdot   D_-  X   \Big) 
  +\ldots
  ~.
\ee
In fact, the following relations hold:
\bea
\int d^2x\,d\theta^+ d\theta^- d\bar \theta^+ d\bar \theta^- (X\bar X-Y \bar Y)
&=&
\int d^2x\,d\theta^+ d\bar \theta^- \bar D_+  D_- (X\bar X-Y \bar Y)~,
\non\\
&=&\int d^2x\,d\theta^+ d\bar \theta^-  \Big(  D_+   X \cdot  \bar  D_-  \bar X - \bar  Y  D_+ \bar  D_-  Y  \Big) 
~.~~~~~~
\eea
Alternatively, 
\bea
\int d^2x\,d\theta^+ d\theta^- d\bar \theta^+ d\bar \theta^- (X\bar X-Y \bar Y)
&=&
\int d^2x\,d \bar \theta^+ d \theta^-   D_+ \bar  D_- (X\bar X-Y \bar Y)~,
\non\\
&=&
\int d^2x\,d \bar \theta^+ d \theta^-  \Big(\bar D_+  \bar X \cdot  D_- X - Y\bar   D_+  D_- \bar Y  \Big) 
~.~~~~~~
\eea
These relations imply
\be 
4 \int d^2x\,d\theta^+ d\theta^- d\bar \theta^+ d\bar \theta^- (X\bar X-Y \bar Y) 
=\int d^2x\,d\theta^+ d\bar \theta^- d \eta^+d  \eta^-   \ { {\bm{\mathcal Y}} } ^2 +c.c.~.
\ee
Once again the  $(4,4)$  supersymmetry of the left hand side becomes manifest on the right hand side. 

As in the $(4,4)$ twisted multiplet considered in the previous subsection, 
we can deform this representation 
to induce the partial breaking.
The deformed multiplet is described by the following $(4,4)$ superfield:
\be
  {\bm{\mathcal Y}}_{\text{def}} =Y +   \eta^+ \bar  D_+ \bar X  +   \eta^-   D_-  X  +   \eta^+  \eta^-   \Big( \bar D_+  D_- \bar Y
  +\kappa\Big) 
  ~.
\ee
The hidden supersymmetry transformations of the component $(2,2)$ superfields can be straightforwardly computed using the arguments of the previous subsection. For the goal of this section, it is enough to mention that
$\d Y$ is the same as the undeformed case of eq.~\eqref{dY2}.

To eliminate half of the degrees of freedom of $  {\bm{\mathcal Y}}_{\text{def}} $ 
and describe a Goldstone multiplet for $\cN=(4,4)\to\cN=(2,2)$ partial supersymmetry breaking,
we again impose the nilpotent constraint
\be
{\bm{\mathcal Y}}^2_{\text{def}}=0 =2  \eta^+ \eta^- \Big(  Y  (\kappa+ \bar D_+  D_- \bar Y) - \bar  D_+ \bar X   \cdot   D_-  X   \Big) +\ldots
~.
\ee
This yields the following constraint for the $(2,2)$ superfields
\be
Y  (\kappa+ \bar D_+  D_- \bar Y) - \bar  D_+ \bar X   \cdot   D_-  X=0
~,
\label{BG44-2}
\ee
which is equivalent to
\be
Y=\frac{ \bar  D_+ \bar X   \cdot   D_-  X}{\kappa+ \bar D_+  D_- \bar Y }
=\frac{  \widetilde W^2  }{\kappa+ \overline{\widetilde  D}^2 \bar Y }
~,\qquad
\bar Y=\frac{ \bar  D_-  \bar X   \cdot   D_+  X}{\kappa+ \bar D_+  D_- \bar Y }
=\frac{ \bar{ \widetilde W}^2  }{\kappa+  \widetilde{   D}^2 \bar Y }
~.
\ee
Here $\bar{\widetilde  D}^2 =\bar  D_+  D_-,  \widetilde D^2 =-  D_+  \bar D_-$ and we have  introduced
the following bilinears:
\be
\widetilde W^2 \equiv   \bar  D_+ \bar X   \cdot   D_-  X =\bar{\widetilde  D}^2  (X\bar X) , \qquad
\bar{ \widetilde W}^2   \equiv  \bar  D_-  \bar X   \cdot   D_+  X  = \widetilde{   D}^2 (X\bar X)
~.
\ee
 
Using exactly the same tricks as before 
and inspired by the $D=4$ Bagger-Galperin model, 
 we can solve the constraints \eqref{BG44-2} to find
\be
Y=\frac{1}{\kappa}\bar{\widetilde  D}^2 \widetilde \Upsilon
~, \quad 
\bar  Y=\frac{1}{\kappa} \widetilde D^2 \widetilde  \Upsilon
~, \qquad 
\widetilde \Upsilon=\bar{\widetilde\Upsilon}= X\bar X - \frac{2 \widetilde W^2 \bar { \widetilde W}^2}{ \widetilde  A+\kappa^2 +\sqrt{ \widetilde  B^2 + 2\kappa^2 \widetilde  A+\kappa^4}   }  
~,
\ee
where 
\bsubea
 \widetilde A&=&\widetilde D^2  \widetilde W^2 +\bar{ \widetilde   D^2} \bar{ \widetilde  W^2}
 =\{ \widetilde   D^2, \bar{ \widetilde   D^2} \}(X\bar X)
 = \partial_{++}  X \partial_{--} \bar{X} + \partial_{++} \bar{X} \partial_{--}  X 
 ~ , 
 \\
 \widetilde  B&=&
  \widetilde  D^2  \widetilde W^2 -\bar{ \widetilde   D}^2 \bar{ \widetilde  W}^2= [ \widetilde   D^2, \bar{ \widetilde   D^2 } ](X\bar X)
  =\partial_{++}  X \partial_{--} \bar{X} - \partial_{++} \bar{X} \partial_{--}  X 
 ~.
\esubea

Since $Y$ is twisted-chiral under the manifest $(2,2)$ supersymmetry \eqref{XYchiral-2}, we can consider the 
twisted-chiral  integral
\be
S_{\kappa^2}=\frac12 \kappa\int d^2x\,d \theta^+ d  \bar \theta^- Y+c.c
=\frac12  \int d^2x\, d   \theta^+ d \bar \theta^-  \bar{\widetilde  D}^2 \widetilde \Upsilon  +c.c.
=\int d^2x\,d\theta^+ d\theta^- d \bar \theta^+ d\bar \theta^-   \widetilde \Upsilon  
 ~.~~~~~~~~~
 \label{chiralYaction}
\ee
By using arguments analogous to those around eqs.~\eqref{deriv-susy-1} of the previous subsection, 
the action \eqref{chiralYaction} proves to be $\cN=(4,4)$ supersymmetric.

Explicitly, the action reads 
\bea
S_{\kappa^2}
= \int d^2x\,d\theta^+ d\theta^- d \bar \theta^+ d\bar \theta^-  
\Bigg( X\bar X - \frac{2 \widetilde W^2 \bar { \widetilde W}^2}{ \kappa^2+\widetilde  A 
+\sqrt{\kappa^4+ 2\kappa^2 \widetilde  A+ \widetilde  B^2 }   }  \Bigg)  
~,
\eea
which precisely matches the model of eq.~\eqref{bgc_on_shell} if we 
identify the coupling constants:
\be
\l =-\frac{2}{\kappa^2}
~.
\ee
This shows explicitly that the $T\Tb$ deformation of the free chiral action possesses a non-linearly realized 
$\cN=(2,2)$  
supersymmetry.


\section{\texorpdfstring{$D=4$}{D=4} \texorpdfstring{$T^2$}{T^2} Deformations and Their Supersymmetric Extensions}\label{Tsquaredcomments} 
 
In section~\C{2D} we exhibited the non-linear supersymmetry possessed by two $D=2$ $\cN=(2,2)$ 
models constructed in \cite{Chang:2019kiu} from the $T\Tb$ deformation of free actions. The striking relationship with the $D=4$ supersymmetric 
Born-Infeld (BI) theory naturally makes one wonder whether some kind of $T\Tb$ flow equation is satisfied by 
supersymmetric $D=4$ BI, 
and related actions. We will spend the rest of the paper exploring this possibility. In this section, we start with a few general 
observations on $T^2$ or supercurrent-squared operators in $D>2$.

 \subsection{Comments on the \texorpdfstring{$T^2$}{T^2} operator in \texorpdfstring{$D=4$}{D=4}}

 In two dimensions, by $T\bar T$ we mean the operator 
 $T_{\mu\nu}T^{\mu\nu}- (T_ \mu^\mu )^2 $, which is proportional to $\det [T_{\mu\nu}]$~\cite{Zamolodchikov:2004ce,Smirnov:2016lqw,Cavaglia:2016oda}. 
 One can attempt to generalize this structure to $D>2$. 
 In general, one could consider the following  stress-tensor squared operator
 \be
 O_{T^2}^{[r]} = T^{\mu\nu}T_{\mu\nu}- r\,\Theta^2 
 ~,
 \qquad
  \Theta\equiv T_ \mu^\mu 
 ~,
 \label{OT2r0}
 \ee  
with $r$ a real constant parameter.
 In two dimensions,  the  unique choice $r=1$ yields a well defined operator which  is free of short distance singularities~\cite{Smirnov:2016lqw,Zamolodchikov:2004ce}. 
 However, to the best of our knowledge, there is no analogous argument in higher dimensions that guarantees a well-defined irrelevant operator $O_{T^2}^{[r]}$ at the quantum level. 
 Nevertheless,  in a $D$-dimensional space-time, one possible extension is given by 
 $O_{T^2}^{[r]}$ with  $r=1/(D-1)$, which  reduces to the $T\bar T$ operator in two dimensions. 
 
 This operator has received some attention recently since it is motivated by a particular holographic picture in $D>2$
\cite{Taylor:2018xcy,Hartman:2018tkw}.
We will not enter into a detailed discussion of the physical properties enjoyed by  $O_{T^2}^{[1/(D-1)]}$, but simply comment that this combination is invariant under a set of improvement transformations of the stress-energy tensor.
Indeed it is easy to show that such a $T^2$ operator transforms by,
 \be
  O_{T^2}^{[1/(D-1)]}  \rightarrow  O_{T^2}^{[1/(D-1)]} +\text{total derivatives}
  ~,
 \ee
 if the (symmetric) stress-energy tensor shifts by the following improvement transformation,
  \be
 T_{\mu\nu} \rightarrow   T_{\mu\nu} +\big(\p_\mu \p_\nu-\eta_{\mu\nu}\p^2 \big)u
 ~,
  \ee
 for an arbitrary scalar field $u$.

In four dimensions, there is another choice of interest, specifically $r=1/2$. 
In fact, it was shown in \cite{Conti:2018jho} that the bosonic Born-Infeld action can be obtained by deforming the free Maxwell theory
with  the operator  $O_{T^2}^{[1/2]}$.\footnote{It is worth mentioning that another type of higher-dimensional generalization 
of $T\Tb$-deformations, specifically the operator $|\det T |^{1/(D-1)}$, was studied in \cite{Cardy:2018sdv,Bonelli:2018kik}.}
In this work, we are going to use $O_{T^2}^{[1/2]}$ as our deforming operator. Once generalized to the
supersymmetric case, we will see that this operator plays a central role for various models possessing non-linearly realized symmetries. 

One interesting property enjoyed by $O_{T^2}^{[1/2]}$ is its invariance under a shift of the Lagrangian density of the theory, or equivalently a shift of the zero point energy. This can serve as motivation for this particular combination. Under a constant  shift  of the Lagrangian 
density $\cL$, and correspondingly its stress-energy tensor,
\be
\cL \rightarrow \cL+ c
~, \qquad 
T^{\mu\nu} \rightarrow T^{\mu\nu}-c\,\eta^{\mu\nu}
~,
\ee
the composite operator $O_{T^2}^{[r]}$ transforms in the following way:
\be
 O_{T^2}^{[r]}  \rightarrow  O_{T^2}^{[r]} +2c(2r-1) \Theta+4c^2(1-r)
 ~.
\ee 
When the theory is not conformal, which is the general situation at an arbitrary point in the flow since the deformation introduces a scale, 
and $r\neq 1/2$, the operator $O_{T^2}^{[r]}$ always transforms in a non-trivial way because of the extra trace term. 
This implies that under a constant shift in the Lagrangian, the dynamics is modified which is certainly peculiar since the shift is trivial in the undeformed theory.\footnote{It is worth noting that $T\bar{T}$ in $D=2$ shares this peculiarity.}

 However if $r=\frac12$,  $O_{T^2}^{[r]}$ is unaffected up to an honest field-independent cosmological constant term. The  shift of the  
 vacuum energy does not affect the dynamics of the theory, as long as the theory is not coupled to gravity. This property is 
 especially interesting, since the $D=4$ $\cN=1$ Goldstino action, which we will study in section  \ref{Goldstino-Flow}
 in the context of $T^2$ flows,
 is the low-energy description of supersymmetry breaking which can generate a cosmological 
 constant. 
For these reasons, we will study the particular operator quadratic in stress-energy tensors given by 
\be\label{Tsquare}
 O_{T^2}\equiv T^{\mu\nu}T_{\mu\nu}- \frac12\Theta^2 
 ~,
\ee 
in the remainder of the paper.

\subsection{\texorpdfstring{$D=4$}{D=4} \texorpdfstring{$\cN=1$}{N=1} supercurrent-squared operator}

We would like to find the $\cN=1$ supersymmetric extension of the $ O_{T^2}$ operator in four dimensions. 
As reviewed in section \ref{2D},
in two dimensions the manifestly supersymmetric $T\bar T$ deformation is roughly given by the square of the supercurrent
superfields.
 One might suspect that a similar construction holds in four dimensions.

 For the remainder of this work, we will assume that the $D=4$ $\cN=1$ supersymmetric theories under our consideration admit
 a Ferrara-Zumino (FZ)  multiplet of currents \cite{Ferrara:1974pz}. Generalizations of this case involving
 the supercurrent multiplets described in
  \cite{Komargodski:2010rb,Dumitrescu:2011iu,Gates:1981yc,Ambrosetti:2016ieg,Gates:1983nr,Dienes:2009td,Kuzenko:2010am}\ 
  might be possible, but merit separate investigation. 
 The operator content of the FZ multiplet, which has 12+12 component fields,
 includes the conserved supersymmetry current $S_{\mu\alpha}$, its conjugate $\bar S_\mu{}^\ad$ 
 and the conserved symmetric energy-momentum tensor $T_{\mu\nu}$:
 \be
 T_{\mu\nu}=T_{\nu\mu}
 ~, \qquad \p^\mu T_{\mu\nu}=0
 ~, \qquad \p^\mu S_\mu = \p^\mu \bar S_\mu=0
 ~.
 \ee
The  FZ multiplet also includes a complex scalar field ${\sf x}$,  as well as 
 the  $R$-current vector field $j_\mu$, which is not necessarily conserved 
 \cite{Ferrara:1974pz}.
 
In $D=4$ $\cN=1$ superspace, 
the FZ multiplet is described by a vector superfield $\cJ_\mu$ and a complex scalar scalar superfield $\cX$ 
satisfying the following constraints:\footnote{We follow the conventions of \cite{Wess:1992cp} 
except for the conversion between vector and bi-spinor indices.
Following \cite{Dumitrescu:2011iu}, we will use the convention
 $ v_{\alpha\dot \alpha}=-2\sigma^\mu_{\alpha\dot \alpha} v_\mu,  \; 
   v_\mu =\frac14 \bar \sigma^{\alpha\dot \alpha}v_{\alpha\dot \alpha}$. 
Then it follows that
 \be
\cJ_{\alpha\dot \alpha}=-2 \sigma^\mu_{\alpha\dot \alpha} \cJ_\mu, \qquad
\cJ^\mu = \frac14 \cJ_{\alpha \dot \alpha} \bar \sigma^\mu{}^{\dot \alpha \alpha}, \qquad
\cJ^2\equiv \eta^{\mu\nu} J_\mu J_\nu=-\frac18  \epsilon^{\alpha \beta}\epsilon^{\dot \alpha \dot \beta}
 \cJ_{\alpha\dot \alpha}J_{\beta\dot \beta}
~.
\ee
}
 \be
 \bar D^{\dot \alpha}  \mathcal J_{\alpha \dot \alpha}=D_\alpha \mathcal X
 ~, \qquad
 \bar D_{\dot \alpha} \mathcal X=0
 ~.
 \ee
The constraints can be solved, and the FZ supercurrents expressed in terms of its $12+12$ 
independent components read\footnote{For convenience, we have rescaled the supersymmetry current compared to 
\cite{Dumitrescu:2011iu}: $ S_\mu^{\text{here}}=-i S_\mu^{\text{there}}$.}
\bea
 \mathcal J_\mu(x)  &=& j_\mu + \theta \Big(S_\mu - { 1 \over \sqrt2} \sigma_\mu \bar \chi\Big) 
 +   \bar\theta \Big(\bar S_\mu + {1  \over \sqrt2} \bar \sigma_\mu \chi\Big) 
 +\frac{i}{2} \theta^2 \p_\mu \bar {\sf x} -  \frac{i}{2} \bar\theta^2 \p_\mu  {\sf x}
\non
\\&&
 + \theta \sigma^\nu \bar \theta 
 \Big( 2 T_{\mu\nu} -\frac23 \eta_{\mu\nu} \Theta -\frac12 \epsilon_{\nu\mu\rho\sigma} \partial^\rho j^\sigma \Big)
 \non\\
 &&
- \frac{i}{2} \theta^2\bar\theta\Big(\bar{\slashed\p}  S_\mu +{ 1\over \sqrt2} \bar \sigma_\mu \slashed\p \bar \chi\Big) 
- \frac{i}{2} \bar\theta^2 \theta\Big( {\slashed\p} { \bar S}_\mu -{1  \over \sqrt2}  \sigma_\mu\bar{  \slashed \p}   \chi \Big) 
\non\\&&
+\frac12 \theta^2 \bar \theta^2 \Big(  \p_\mu\p^\nu j_\nu -\frac12 \p^2 j_\mu \Big) 
~,
\eea
and
\bsubea
\mathcal X(y)&=&{\sf x}(y)  +\sqrt2 \theta \chi(y)   +\theta^2 {\sf F}(y)
~,
~~~
\\
\chi_\alpha
&=& \frac{  \sqrt{2}   }{3}( \sigma^\mu)_{\alpha\dot\alpha}\bar S^{\dot \alpha}_\mu
~,
\qquad
{\sf F}=  \frac23 \Theta +i \partial_\mu j^\mu
  ~,
\esubea
where the chiral coordinate $y^\mu=x^\mu +i \theta\sigma^\mu \bar \theta$,
and  we used $\slashed\p=\sigma^\mu\p_\mu, \, \bar{\slashed\p}=\bar\sigma^\mu\p_\mu$.

If we seek a manifestly supersymmetric completion of the operator \eqref{Tsquare} 
by using combinations of the supercurrent superfields with dimension $4$,
it is clear that the only possibility is the full superspace
integral of a linear combination of $\cJ^2$ and $\cX\bar\cX$.
Up to total derivatives and terms that vanish by using the supercurrent conservation equations, or equivalently that vanish
on-shell, the $D$-terms of $\cJ^2$ and $\cX\bar\cX$ are given by\footnote{The composite $A$ (and analogously its conjugate $\bar{A}$)   is given by
\beqn
A&=&\Big(S_\mu - { 1 \over \sqrt2} \sigma_\mu \bar \chi\Big)
\Big( {\slashed\p} { \bar S}^\mu -{1  \over \sqrt2}  \sigma^\mu\bar{  \slashed \p}   \chi \Big)
=S_\mu {\slashed\p} { \bar S}^\mu  -  \bar \chi  \bar{  \slashed \p}   \chi  +\sqrt2  { \bar S}^\mu \p_\mu \bar\chi
+\text{total derivatives}
~.
\eeqn
The equality can be obtained with some algebra. Note that the last term drops  
after integration by parts because of the conservation equation for $S_\mu$.
}
\bsubea
\mathcal J^2 |_{\theta^2\bar\theta^2} \equiv \eta^{\mu\nu} \cJ_\mu \cJ_\nu|_{\theta^2\bar\theta^2}
&=&
-\frac12  \Big( 2 T_{\mu\nu} -\frac23 \eta_{\mu\nu} \Theta -\frac12 \epsilon_{\nu\mu\rho\sigma} \partial^\rho j^\sigma \Big)
 ^2  +j^\mu \Big(  \p_\mu\p^\nu j_\nu -\frac12 \p^2 j_\mu \Big)  
\non\\
&&
 +\frac12 \p_\mu {\sf x}\p^\mu \bar {\sf { x}}
+\frac{i}{2} \Big(A-\bar A \Big)
\\&=&
-2(T_{\mu\nu})^2   +\frac49 \Theta^2 - \frac54  \Big( \p_  \mu j^\mu \Big)^2 -\frac34 j_\mu \p^2  j^\mu + \frac12 \p_\mu {\sf \bar x}\p^\mu {\sf x}
\non\\
&&
 +i  \Big( S_\mu {\slashed\p} { \bar S}^\mu  -   \bar \chi  \bar{  \slashed \p}   \chi 
 \Big)
  +\text{total derivatives} 
  +{\rm EOM}
  ~,
\esubea
and
\bsubea
\mathcal X \bar\cX |_{\theta^2\bar\theta^2} 
&=&
  {\sf F} \bar   {\sf F} -\p_\mu {\sf x}\p^\mu\bar {\sf x}
-i \bar \chi{ \bar{\slashed \p }}  \chi  +\text{total derivatives} 
\\
&=&   \frac49 \Theta^2 +(\p_\mu j^\mu)^2-\p_\mu {\sf x}\p^\mu\bar {\sf x}
-i \bar \chi{ \bar{\slashed \p }}  \chi
 +\text{total derivatives} 
 ~.
\esubea

To get a manifestly supersymmetric extension of  $O_{T^2}=T^2 -\frac12 \Theta^2$, we have to consider the following linear 
combination 
   \be\label{superTsquare}
  \mathcal O_{T^2}= -\frac{1}{2} \Big( \eta^{\mu\nu} \mathcal J_\mu \mathcal  J_\nu+\frac54 \mathcal X\bar {\mathcal X} \Big)= 
   \frac{1}{16} \mathcal  J^{\alpha\dot\alpha}   \mathcal J_{\alpha\dot\alpha}- \frac58  \mathcal  X    \bar {\mathcal    X}
   ~.
   \ee 
In fact, the supersymmetric descendant of the supercurrent-squared operator $  \mathcal O_{T^2}$ is 
 \bsubeq
 \label{superTsquare2}
 \bea
{\sf  O}_{T^2}&=& \int d^4 \theta \,  \mathcal O_{T^2} 
\\&=&
 T^2 -\frac12\Theta^2 +\frac38 j_\mu \p^2 j^\mu +\frac38 \p_\mu {\sf x}\p^\mu\bar {\sf x}
-\frac{i}{2}  \Big( S_\mu {\slashed\p} { \bar S}^\mu  - \frac94  \bar \chi  \bar{  \slashed \p}   \chi   \Big)
\non\\
&&
 +\text{total derivatives}+{\rm EOM}
~.
 \eea
 \esubeq
This result shows that ${\sf O}_{T^2}$ is the natural supersymmetric extension of $O_{T^2}$. 
However, it is worth emphasizing that in the $D=4$ case,  the supersymmetric descendent ${\sf  O}_{T^2}$ of $\cO_{T^2}$ 
has extra non-trivial contributions from other currents. This should be contrasted with the $D=2$ case where ${\sf  O}_{T^2}=O_{T^2}$
up to ${\rm EOM}$ and total derivatives, see eq.~\eqref{calTTb=TTb}.

It actually does not seem possible to find a linear combination of $\cJ^2$ and $\cX\bar\cX$ such that an analogue 
of eq.~\eqref{calTTb=TTb}  holds in $D=4$.
This suggests that, in contrast with the $D=2$ case, deformations of a Lagrangian triggered
by the operators $O_{T^2}$ and ${\sf  O}_{T^2}$
will in general lead to different flows: one manifestly supersymmetric, while the other not.


 \section{Bosonic Born-Infeld As a \texorpdfstring{$T^2$}{T^2} Flow}
\label{bosonicBI}

It was shown in \cite{Conti:2018jho} that the $D=4$ Born-Infeld action arises from a $D>2$ generalization of the $T\bar T$ deformation.  Specifically, the operator driving the flow equation was shown to be the $ O_{T^2}$ defined in eq.~\eqref{Tsquare} of the preceding section.
In this section we review this result  in detail as it is a primary inspiration for our supersymmetric extensions. 
   
The $D=4$ bosonic BI action on a flat background
 is given by 
\bea
S_{\text{BI }}
&=&    \frac{1}{\alpha^2} \int d^4x\;  \Big[  1-  \sqrt{-\det (\eta_{\mu\nu}+\alpha F_{\mu\nu } )}   \Big]~ 
\non\\
&=&  \frac{1}{\alpha^2}\int d^4x\;  \Big[  1-  \sqrt{1+\frac{\alpha^2}{2}  F^2 -\frac{\alpha^4}{16} (F\tilde F)^2    }   \Big]~ 
\non\\
&=&
 -\frac14\int d^4x\;  F^2
 +{\rm higher~derivative~terms}
~,
\label{flat_BI}
\eea
where $F_{\mu\nu}=(\pa_{\mu}v_\nu-\pa_{\nu}v_\mu)$ is the field strength for an Abelian gauge field $v_\mu$, and
\be
F^2\equiv F_{\mu\nu}F^{\mu\nu}, \qquad   F\tilde F\equiv F_{\mu\nu} \tilde F^{\mu\nu}=\frac12 \epsilon_{\mu\nu\rho\sigma}F^{\mu\nu}F^{\rho\sigma}
~.
\ee

The  stress-energy tensor
  for the BI action can be  computed straightforwardly and it reads  \cite{Rasheed:1997ns} 
\be 
T^{\mu\nu} =-\frac{  F^{\mu   \lambda} F^\nu{}_{\lambda  } +
 \frac{1}{\alpha^2}  \Big(\sqrt{1+\frac{\alpha^2}{2}  F^2 -\frac{\alpha^4}{16} (F\tilde F)^2    } -1-\frac{\alpha^2}{2}  F^2\Big) \eta^{\mu\nu}   }
 {  \sqrt{1+\frac{\alpha^2}{2}  F^2 -\frac{\alpha^4}{16} (F\tilde F)^2    }  }
 ~.
\ee
This can be written in the following useful form
\be
T^{\mu\nu}   =   \frac{  T^{\mu\nu} _{\text{Maxwell}}  } { \sqrt{1+2 A +  B^2    }  } 
+
\frac{  \eta ^{\mu\nu} }{\alpha^2\sqrt{1+2    A +   B^2    }  }
\frac{ A^2 - B^2}{ 1+  A  +\sqrt{1+2    A +  B^2    }   } 
~,
\ee
where we used the stress-energy tensor for the Maxwell theory 
\be
T^{\mu\nu} _{\text{Maxwell}} = - F^{\mu   \lambda} F^\nu{}_{\lambda}  +\frac14 F^2  \eta ^{\mu\nu}
~,
\ee
while $A$ and $B$ are defined by
\be\label{4DAB}
A=\frac14 \alpha^2 F^2
~, \qquad 
B=\frac{i}{4}   \alpha^2 F\tilde F
~.
\ee
 It is easy to compute the trace of the stress-energy tensor 
\be
\Theta=T^{\mu\nu} \eta_{\mu\nu} = \frac{ 4}{\alpha^2\sqrt{1+2    A +   B^2    }  }
\frac{ A^2 - B^2}{ 1+  A  +\sqrt{1+2    A +  B^2    }   } 
~,
\ee
where, interestingly, the combination $(A^2-B^2)$ proves to be related to the square of $T^{\mu\nu}_{\text{Maxwell}}$. Using the identity 
 \be
 (F\tilde F)^2= \frac14 ( \epsilon_{\mu\nu\rho\sigma}F^{\mu\nu}F^{\rho\sigma})^2
 =4F_{\mu\nu}F^{\nu \rho}F_{\rho \sigma}F^{\sigma \mu} -2 (F^2)^2
 ~,
 \ee
we see that 
\be
T^2 _{\text{Maxwell}}=F_{\mu\nu}F^{\nu \rho}F_{\rho \sigma}F^{\sigma \mu} -\frac14 (F^2)^2
=\frac14 \Big(  (F^2)^2 +(F\tilde F)^2 \Big) 
=\frac{4}{\alpha^4}(A^2-B^2)
~.
\ee

Using tracelessness of the free Maxwell stress-energy tensor,
the $O_{T^2}$ operator can be easily computed:
\bsubeq\beqn\label{OT2}
O_{T^2}=T^2 -\frac12 \Theta^2&=&
  \frac{4(  A^2 -B^2)   } {\alpha^4 \sqrt{1+2 A +  B^2    }^2  } 
  \Big( 1
-
\frac{ A^2 - B^2}{ (1+  A  +\sqrt{1+2    A +  B^2    } )^2  }  \Big)~,
~~~~~~~~~
\\&=&
  \frac{4(  A^2 -B^2)   } {\alpha^4 \sqrt{1+2 A +  B^2    }^2  } 
  \Big( 1
-
\frac{ 1+  A  -\sqrt{1+2    A +  B^2    }  }{ 1+  A  +\sqrt{1+2    A +  B^2    }   }\Big)~,
\\&=&
  \frac{8(  A^2 -B^2)   } {\alpha^4 \sqrt{1+2 A +  B^2    }   } 
\frac{ 1      }{ 1+  A  +\sqrt{1+2    A +  B^2    }} ~,
\\&=&
  \frac{8(  1+  A  -\sqrt{1+2    A +  B^2 )   }} {\alpha^2 \sqrt{1+2 A +  B^2    }   } 
  ~.
\eeqn
\esubeq
The variation of the BI Lagrangian with respect to the 
parameter $\a^2$ can be readily computed, 
and it is given by
\be
\frac{\p \mathcal L_\alpha}{\p \alpha^2} = \frac{1+\frac14 \alpha^2 F^2- \sqrt{1+\frac12 \alpha^2 F^2 -\frac{1}{16} \alpha^4 (F\tilde F)^2}}{\alpha^2 \sqrt{1+\frac12 \alpha^4 F^2 -\frac{1}{16} \alpha^4 (F\tilde F)^2}}
~.
\label{der-BI}
\ee
Once we use \eqref{4DAB} it is clear that \eqref{OT2} and \eqref{der-BI} have exactly the same structure and satisfy
the following equivalence equation
\be
\frac{\p\mathcal L_\alpha}{\p \alpha^2} =\frac{1}{8} O_{T^2}
~,
\ee
showing that the BI Lagrangian satisfies a $T^2$-flow driven by the operator $O_{T^2}$.

Before turning to $D=4$ supersymmetric analysis, it is worth mentioning that the structure  
of the computation relating  the $O_{T^2}$ operator to the bosonic BI theory, which we just reviewed,
is quite similar to what we saw in section \ref{2D} for the $D=2$ $\cN=(2,2)$ supersymmetric $T\bar T$ flows. 
For example, in the deformation of the free twisted-chiral multiplet action,  
the analogue of the $A$ and $B$ combinations of \eqref{4DAB} is given by 
\eqref{2DAB}, but the square root structure of the actions is completely analogous. 
This fact, together with the non-linearly realized supersymmetry we investigated in section \ref{2D},
naturally lead to the guess that the $D=4$ $\cN=1$ supersymmetric Born-Infeld (BI) theory 
 may also satisfy a  $T^2$ flow. 
The next section is devoted to explaining how this is the case.


\section{Supersymmetric Born-Infeld From Supercurrent-Squared Deformation}
\label{BI-flows}

In section \ref{2D}
we proved, by analogy and extension of the $D=4$ results of \cite{Bagger:1996wp},  
that two $D=2$ supercurrent-squared flows possess additional non-linearly realized supersymmetry.
In this section we reverse the logic.
We will look at a well-studied model, namely the Bagger-Galperin construction
\cite{Bagger:1996wp} of $D=4$ $\cN=1$ Born-Infeld theory
\cite{Deser:1980ck,cecotti:1987}, and show that it satisfies a supercurrent-squared flow equation.

\subsection{\texorpdfstring{$D=4$}{D=4} \texorpdfstring{$\cN=1$}{N=1} supersymmetric BI  and non-linear supersymmetry}

Let us review some well known results about the $D=4$ $\cN=1$ Born-Infeld theory
\cite{cecotti:1987}, the 
Bagger-Galperin action \cite{Bagger:1996wp}, the non-linearly realized second supersymmetry, and its precise $\cN=2\to\cN=1$
supersymmetry breaking pattern. 
For more detail, we refer to the following references on the subject \cite{Bagger:1996wp,Rocek:1997hi,Kuzenko:2015rfx,Antoniadis:2017jsk,Antoniadis:2019gbd,cecotti:1987}.

We start  with the following  $\mathcal N=2$ superfield,
\be\label{n2chiral}
\mathcal  W (y, \theta, \tilde \theta)= X(y,\theta )+\sqrt{2} i \tilde  \theta  W(y,\theta ) - \tilde \theta^2 G(y,\theta )
~, 
\qquad y^\mu=x^\mu+ i \ta \sigma ^\mu \tab+ i \tilde \ta \sigma ^\mu \bar{\tilde\theta}
~,
\ee
which is chiral with respect to both supersymmetries:\footnote{We follow the conventions of \cite{Wess:1992cp}. 
The $D=4$, $\cN=2$ superspace is parametrised by bosonic coordinates $x^\mu$ and the 
Grasmannian coordinates $(\theta^\a,\,\bar{\theta}^\ad)$ and $(\tilde \theta^\a,\,\bar{\tilde \theta}^\ad)$.
In terms of the chiral coordinate $y^\mu$ introduced in \eqref{n2chiral}, 
the supercovariant {derivatives} are given by 
\be
D_\alpha =\frac{\p}{\p \ta^\alpha}+2i \sigma^\mu_{\alpha \dot \alpha}\tab^{\dot \alpha} \frac{\p}{\p y^\mu},\qquad \bar D_{\dot\alpha} =-\frac{\p}{\p \bar\ta^{\dot\alpha}}
~,
\ee
and similarly for $\tilde D_\alpha,  \bar{ \tilde{D}}_{\ad}$.
}
\be
\bar D_\ad\mathcal W= \bar{ \tilde{D}}_\ad\mathcal W=0~.
\ee
Since we are ultimately interested in partial $\cN=2\to\cN=1$ supersymmetry breaking, 
we will mostly use $\cN=1$ superfields associated to the $\q$ Grassmann variables to describe manifest supersymmetry,
while we use the $\tilde{\q}$ variable for the hidden non-linearly realized supersymmetry.
 The $\cN=1$ superfields $X$, $W_\a$, and $G$ of eq.~\eqref{n2chiral} 
  are chiral under the manifest $\cN=1$ supersymmetry. Under the additional hidden $\cN=1$ supersymmetry, they transform as follows:
\bsubeq\beqn
\tilde \delta X&=&\sqrt2 i \epsilon W~, \\
\tilde \delta W &=&\sqrt2  \sigma^\mu \bar \epsilon  \p_\mu X+\sqrt2  i \epsilon G~, \\
\tilde \delta  G&=& - \sqrt2 \p_\mu W \sigma^\mu \bar \epsilon~.   \label{susytsfG}
\eeqn\esubeq

The superfield \eqref{n2chiral} has 16+16 independent off-shell components and is reducible. 
It contains the degrees of freedom of an $\mathcal N=2$ vector and tensor multiplet.
To reduce the degrees of freedom and describe an irreducible $\cN=2$ off-shell vector multiplet,
we impose the following conditions on the $\cN=1$ components of $\cal W$:
\begin{itemize}
\item[{(i)}] First that $W_\a$ is the field-strength superfield of an $\mathcal N=1$ vector multiplet satisfying,
\be
D^\a W_\a-\bar D_\ad \bar W^\ad=0
~,
\ee
\item[{(ii)}] 
and that
\be
G=\frac14 \bar D^2 \bar X
~.
\label{constrW2}
\ee
\end{itemize}
The latter condition can easily be seen to be consistent since
it is straightforward to verify 
that $\frac14 \bar D^2 \bar X$ transforms in the same way  {as  $G$} given in \eqref{susytsfG}. 
Therefore 
we can impose \eqref{constrW2}
without violating $\mathcal N=2$  supersymmetry.

Since $\mathcal W$ is chiral with respect to both sets of supersymmetries, we can consider the following  Lagrangian, 
\bea
\mathcal L^{\mathcal N=2}_{ \mathcal W^2} 
&=& 
\frac14\int d^2 \theta  d^2 \tilde \theta \,\mathcal W^2+c.c.
=\frac14 \int d^2 \theta \Big(W^2-\frac12 X\bar D^2 \bar X\Big)+c.c.~.
\eea
On the other hand,  the $\mathcal N=2$  Maxwell theory written in terms of the $\mathcal N=1$ chiral superfields $X$ and $W_\a$ is  given by 
\bea\label{N2Maxwell}
\mathcal L^{\mathcal N=2}_{ \text{ Maxwell}}
&=&
\int d^2 \theta d^2 \bar \theta\, \bar X X +\frac14 \int d^2 \theta \, W^2
+\frac14 \int d^2 \theta\,  \bar W^2~,
\non\\
&=&\frac14 \int d^2 \theta \Big(W^2-\frac12 X\bar D^2 \bar X\Big)+c.c.
+{\rm total~derivative}
~.
\eea
We see that these two Lagrangians are the same, confirming that the extra constraint imposed on $\mathcal W$ is correct.  
 The  off-shell $\mathcal N=2$  vector multiplet can therefore be described in term of the following $\mathcal N=2$ superfield
\be\label{n2vector}
\mathcal  W (y, \theta, \tilde \theta)= X(y,\theta )+\sqrt{2} i \tilde  \theta  W(y,\theta ) -\frac14 \tilde \theta^2  \bar D^2 \bar X(y,\theta)~,
\ee
where $X$ and $W_\a$ are  $\mathcal N=1$ chiral and vector  multiplets, respectively. {Their component expansion reads:}
\bsubeq\beqn
W_\alpha &=& -i \lambda_\alpha +  \theta_\alpha \sfD  -  i (\sigma^{\mu\nu}\theta)_\alpha F_{\mu\nu}    +\q^2 (\sigma^\mu   \p_\mu {\bar \lambda}  )_\alpha~,
\label{N1_vector_components}
\\
X&=&x+\sqrt{2} \theta \chi -\theta^2\, \sfF~.
\eeqn
 \esubeq
 
Following    \cite{Rocek:1997hi} (see also \cite{Kuzenko:2015rfx,Antoniadis:2017jsk,Antoniadis:2019gbd}), 
we break  $\mathcal N=2$ supersymmetry by considering a Lorentz  and $\cN=1$  invariant condensate with a non-trivial dependence on the hidden Grassmann variables   $\EV{\mathcal W}=\mathcal W_{\text{def}} \propto \tilde \theta^2  \neq 0$, 
such that
\bsubea
\mathcal W&\rightarrow&  {\mathcal W_{\text{new}}}=\EV{\mathcal W}+\mathcal W= \mathcal W+\mathcal W_{\text{def}}~,
\\
{\mathcal W_{\text{new}}}
&=&
X +\sqrt{2} i \tilde  \theta  W  
 -\frac14 \tilde \theta^2 \Big(   \bar D^2 \bar X  + \frac{2}{ \kappa}\Big)
 ~.
 \label{Wnew}
\esubea
The hidden supersymmetry transformations of the $\cN=1$ components of the deformed $\cN=2$ vector multiplet turn out to be
\bsubea
\tilde \delta X
&=&
\sqrt2 i \epsilon W
\label{hidden_susy-1}
~, \\
\tilde \delta W 
&=&
\frac{i}{\sqrt2 \kappa}\epsilon 
+\frac{i}{2\sqrt2}   \epsilon \Db^2\bar{X}
+\sqrt2  \sigma^\mu \bar \epsilon  \p_\mu X
~.
\label{hidden_susy-2}
\esubea
Assuming the model under consideration preserves the manifest $\cN=1$ supersymmetry, which implies $\EV{\Db^2X}=0$,
the explicit non-linear $\kappa$-dependent term in the transformation of the fermionic $W_\a$ signals 
the spontaneous partial breaking $\cN=2\to\cN=1$ of the hidden supersymmetry.

To describe the Maxwell-Goldstone multiplet for the partial breaking $\cN=2\to\cN=1$, we can impose the following nilpotent constraint on the deformed $\cN=2$ superfield strength ${\mathcal W_{\text{new}}}$
\cite{Rocek:1997hi}
\be\label{N2constraint}
( {\mathcal W_{\text{new}}})^2=0~.
\ee
Once reduced to $\cN=1$ superfields, following the expansion \eqref{Wnew}, this constraint implies
the Bagger-Galperin constraint \cite{Bagger:1996wp}
\be\label{XWeq}
\frac1\kappa X=W^2-\frac12  X\bar D^2 \bar X
~,
\ee 
which can be solved to eliminate $X$ in terms of 
$W^2=W^\a W_\a$ and its complex conjugate 
$\bar{W}^2=\bar{W}_\ad \bar{W}^\ad$:
 \be\label{XWconstraint}
 X=\kappa W^2 -\kappa^3 \bar D^2 \Big[   \frac{W^2 \bar W^2}{ 1+ \mathcal A +\sqrt{1+2 \mathcal A- \mathcal B^2}}  \Big]~,
 \ee
 where we have introduced:
 \be\label{AtBt}
 \mathcal A=\frac{\kappa^2}{2} (D^2 W^2+\bar D^2 \bar W^2 )= \overline{\mathcal A}
 ~, \qquad 
 \mathcal B= \frac{\kappa^2}{2} (D^2 W^2-\bar D^2 \bar W^2 )= - \overline{\mathcal B}~.
 \ee
 For later use we denote the lowest components of the composite superfields $\cal A$ and $\cal B$  
 \be
 A= \mathcal A |_{\theta=0}
 ~, \qquad  
 B= \mathcal B |_{\theta=0}~.
 \ee
We will not repeat the derivation of \eqref{XWconstraint} which can be found in the original paper
 \cite{Bagger:1996wp}, and was reviewed and slightly modified in section \ref{2D} for our analysis in two dimensions.

 The $\cN=1$ supersymmetric BI action can be constructed using the following $\cN=1$ (anti-)chiral Lagrangian linear in $X$:
\be\label{BI}
 \mathcal L_\kappa=\frac{1}{4 \kappa   } \Big(\int d^2 \ta X+\int d^2 \tab \bar X \Big)~.
 \ee
The second hidden supersymmetry eq.~\eqref{hidden_susy-1} written in terms of the 
unconstrained real vector multiplet $V$, where $W_\a=-1/4 \Db^2D_\a V$, takes the form:
\be
\tilde \delta X=-\frac14 \sqrt2 i \epsilon^\alpha    \bar D^2 D_\alpha V
~.
\ee
Using the fact that $D^2\Db^2D_\a\propto\pa_{\a\ad}D^2\Db^\ad$,
one can immediately see that the supersymmetry variation of $\mathcal L_\kappa $ in \eqref{BI} is a total derivative. 
Therefore this supersymmetric BI action is invariant under the second hidden non-linear supersymmetry. 
  
Using the solution \eqref{XWconstraint}, the  supersymmetric BI Lagrangian  takes the explicit form 
  \beqn
 \mathcal L_\kappa&=& 
 \frac{1}{4\kappa}\int d^2 \theta \Big(\kappa W^2 -\kappa^3 \bar D^2 \Big[   \frac{W^2 \bar W^2}{ 1+ \mathcal A +\sqrt{1+2 \mathcal A+ \mathcal B^2}}  \Big] \Big)
 +c.c.~,
 \non
\\
 &=& \frac14 \int d^2 \theta \, W^2+ \frac14 \int d^2\bar \theta \, \bar W^2 
 +2 \kappa^2 \int d^2 \theta d^2 \bar\theta  \,\frac{W^2 \bar W^2}{ 1+ \mathcal A +\sqrt{1+2 \mathcal A+ \mathcal B^2}}  
 ~,
 \label{susyBI1}
  \eeqn
which makes it clear that  the supersymmetric BI is a non-linear deformation of the free $\cN=1$ Maxwell theory.  
  This supersymmetric extension of BI was first constructed by Bagger and Galperin in
\cite{Bagger:1996wp}. In this work when we refer to the supersymmetric BI theory, we will always
mean the Bagger-Galperin action.
  
We can easily calculate the flow under the $\kappa^2$ coupling constant,
\be\label{BIflow}
\frac{\p \mathcal L_\kappa}{\p \kappa^2}=2 \int d^2 \theta d^2\bar \theta\,
\frac{  W^2 \bar W^2   }{ 1+\mathcal A  +\sqrt{ 1+2 \mathcal A+\mathcal  B^2    }   } \frac{1}{\sqrt{ 1+2 \mathcal A+\mathcal  B^2    } }
~.
\ee
Our goal is now to show that the right hand side of this flow equation on-shell
is the specific supercurrent bilinear \eqref{superTsquare}  that we introduced earlier. This will establish a supercurrent-squared flow for the supersymmetric BI action.

Before turning to the core of this analysis
let us recall that at the leading order in $\kappa^2$, the fact that $D=4$ $\cN=1$ BI 
satisfies a supercurrent-squared flow was already noticed in \cite{cecotti:1987}.
This result was also highlighted recently in the introduction of \cite{Chang:2018dge}.
In fact, note that in the free limit $\alpha =\kappa=0$,  the Lagrangian \eqref{susyBI1} becomes 
the $\cN=1$ supersymmetric Maxwell theory. 
Its supercurrent multiplet is
\be
\cJ_{\alpha \dot \alpha}=-4W_\alpha \bar W_{\dot \alpha}
~, \qquad  
\cX=0
~,
\ee
where $\cX=0$ because super-Maxwell theory is scale invariant.  
The supersymmetric $T^2$ deformation operator  \eqref{superTsquare} is then simply given by
\be
\mathcal O_{T^2}=   \frac{1}{16}   \cJ_{\alpha\dot \alpha}\cJ^{\alpha\dot \alpha}  -\frac58 \cX \bar \cX=  W^2 \bar W^2
~,
\ee 
and to leading order  \eqref{BIflow} turns into \cite{cecotti:1987}
\be
\frac{\p \mathcal L_\kappa}{\p \kappa^2}=\int d^2 \theta d^2\bar \theta\,  W^2 \bar W^2 +\mathcal O(\kappa^2)
=\int d^2 \theta d^2\bar \theta\,   \mathcal O_{T^2}+\mathcal O(\kappa^2)
~.
\ee
This shows that the supercurrent-squared flow equation 
is satisfied at this order.
The rest of this section is
devoted to demonstrating the  full non-linear
extension of this result.
First, we are going to look at the bosonic truncation of \eqref{susyBI1} and \eqref{BIflow}.

\subsection{Bosonic truncation}
\label{bosonicBI-2}

  In the pure bosonic case
  the gauginos are set to zero in \eqref{N1_vector_components}, $\lambda=\bar \lambda=0$, and  $W^2, \bar W^2$ only have $\theta^2, \bar \theta^2$ components, so $\mathcal A, \mathcal B$ can only contribute the  lowest components:
\be
 A=\mathcal A|_{\theta=0}= 2 \kappa^2  \Big( F^2 - 2 \sfD^2 \Big)
 ~,\qquad
 B=\mathcal  B|_{\theta=0}=  2    \kappa^2  i F \tilde F
 ~.
\ee
Therefore the supersymmetric BI Lagrangian reduces to 
 \be
 \mathcal L=  
 \frac{1}{8\kappa^2} \Big[1- \sqrt{1+4 \kappa^2  \Big( F^2 - 2\sfD^2 \Big) -4  \kappa^4  \Big( F \tilde F  \Big)^2
}   \Big] 
~.
 \ee
The auxiliary field $\sfD=0$ after using its ${\rm EOM}$, 
and the Lagrangian is equivalent to the bosonic BI Lagrangian \eqref{flat_BI} with the identification $\alpha^2=8 \kappa^2$. 
This immediately implies that on-shell the bosonic truncation of the supersymmetric BI
satisfies a $T^2$ flow equation driven by the $O_{T^2}$ operator \eqref{Tsquare}, as we discussed in \eqref{OT2}.
A similar story is going to hold for the complete supersymmetric model of \eqref{susyBI1} and \eqref{BIflow}.

\subsection{Supersymmetric Born-Infeld as a supercurrent-squared flow}

The supercurrent for the supersymmetric BI action 
\eqref{susyBI1} was computed in \cite{Kuzenko:2002vk}  for  $\kappa^2=\frac12$. 
To simplify notation, we will also consider the special case  $\kappa^2=\frac12$ in our intermediate computations.  
The $\kappa$-dependence can be restored easily and will appear in the final formulae. 

We can straightforwardly use the results of \cite{Kuzenko:2002vk} for our supercurrent-squared flow analysis. 
The FZ multiplet was computed for a class of 
models described by the following Lagrangian,
\bea
\cL 
&=& 
\frac14 \int d^2 \theta \, W^2
+ \frac14 \int d^2\bar \theta \, \bar W^2 
+\frac{1}{4} \int d^2 \theta d^2 \bar\theta  \, W^2 \bar W^2\L(u,\bar{u})
 ~,
 \label{GeneralBI}
\eea
where
\be
u=\frac18 D^2 W^2
~, \qquad \bar u=\frac18 \bar D^2 \bar W^2
~.
\ee
The action \eqref{susyBI1} turns out to be given by the following choice of $\L(u,\bar{u})$
\be
\Lambda(u,\bar u)=\frac{4}{  1+ \mathcal A +\sqrt{1+2 \mathcal A+ \mathcal B^2}  }
~,
\label{LambdaBI}
\ee
where
\be
\mathcal A=2(u+\bar u)
~ , \qquad
\mathcal B=2(u-\bar u)
~.
\ee 
Following \cite{Kuzenko:2002vk}, we also introduce the composite superfields
\be
\Gamma(u,\bar u)=\frac{\p(u\Lambda)}{\p u}
~, \qquad
\bar \Gamma(u,\bar u)=\frac{\p(\bar u\Lambda)}{\p \bar u}
~,
\ee
which, in the case of interest to us where \eqref{LambdaBI} holds, satisfy
\bsubeq
\beqn \label{gammau}
\Gamma +\bar \Gamma -\Lambda&=&\frac{4}{\Big( 1+ \mathcal A +\sqrt{1+2 \mathcal A+ \mathcal B^2} \Big) \sqrt{1+2 \mathcal A+ \mathcal B^2} }
~,
\\
\bar u \Gamma+ u \bar \Gamma
&=& 1 -\frac{1}{  \sqrt{1+2 \mathcal A+ \mathcal B^2}  }  
~.
\eeqn
\esubeq
The supercurrents will also be functionals of the following composite
\bsubeq
\beqn
i M_\alpha &=&W_\alpha \Bigg[ 1-\frac14 \bar D^2 \Bigg(\bar W^2\Big(\Lambda +\frac18 D^2(W^2 \frac{\p \Lambda}{\p u})\Big)  \Bigg)   \Bigg]~,  \\
&=&
W_\alpha \Big( 1- 2 \bar u\Gamma \Big)+  W \bar W (\cdots)+   W^2 (\cdots)
~,
\eeqn
\esubeq
where $W \bar W (\cdots)$ denotes terms which are proportional to $W_\alpha  \bar W_{\dot \alpha}$, 
while $  W^2 (\cdots)$ denotes terms  proportional to $W^2$.
We will use similar notation with ellipses  denoting quantities  with bare fermionic terms that will not contribute to the calculation because of nilpotency conditions.

 With the ingredients introduced above, the FZ multiplet for 
 the supersymmetric BI action is given by  
\cite{Kuzenko:2002vk} 
\bsubeq
\beqn
\mathcal X&=&  \frac16 W^2 \bar D^2 \Big( \bar W^2(\Gamma+\bar \Gamma -\Lambda  ) \Big) 
~,
\\
\mathcal J_{\alpha \dot{\alpha}} &=&  -2 i M_\alpha \bar W_{\dot \alpha}+2i W_\alpha \bar M_{\dot \alpha}
+\frac{1}{12} [ D_\alpha, \bar  D_{\dot \alpha}]  \Big(W^2 \bar W^2 \Big) \cdot \Big(\Gamma +\bar \Gamma -\Lambda \Big)
\non\\
&&
 +W^2 \bar W  (\cdots)+ \bar  W^2  W (\cdots)
 ~.
\eeqn
\esubeq
For our purposes, the superfields $X$ and $\cJ_{\a\ad}$ can be further simplified as follows:
\bsubeq
\bea
\mathcal X&=&\frac16  W^2 \bar D^2  \bar W^2\cdot  \Big(     \Gamma+\bar \Gamma -\Lambda  \Big) +W^2 \bar W (\cdots)~,
\\
&=&
\frac{2W^2 \bar D^2  \bar W^2}{3 \Big( 1+\mathcal A  +\sqrt{ 1+2 \mathcal A+\mathcal  B^2    } \Big)  }+W^2 \bar W (\cdots)
~,
\label{X2}
\eea
\esubeq
and
\bsubeq
\bea
\mathcal J_{\alpha \dot{\alpha}} &= &
 -4W_\alpha \bar W_{\dot \alpha} (1- \bar u \Gamma  -   u\bar\Gamma)
+\frac{1}{12 } [ D_\alpha, \bar D_{\dot \alpha}]  \Big(W^2 \bar W^2 \Big) \cdot \Big(\Gamma +\bar \Gamma -\Lambda \Big)
\non\\
&&
 +W^2 \bar W  (\cdots)+ \bar  W^2  W(\cdots) ~,
\\&= & -\frac{ 4W_\alpha \bar W_{\dot \alpha}  }{\sqrt{1+2 \mathcal A+ \mathcal B^2}}
+\frac{ 2 D_\alpha W^2  \cdot \bar  D_{\dot \alpha}  \bar W^2  }{3\Big( 1+ \mathcal A +\sqrt{1+2 \mathcal A+ \mathcal B^2} \Big) \sqrt{1+2 \mathcal A+ \mathcal B^2} }
\non\\
&&
 +W^2 \bar W  (\cdots)+ \bar  W^2  W (\cdots)
 ~,
\label{trace}
\eea
\esubeq
where we used \eqref{gammau}. 

The computation of $\mathcal  X \bar \cX $ is trivial and receives contributions only from the square of the first term in \eqref{X2}.
The computation of $\mathcal J^2$ is less trivial. 
It is obvious that the last two complicated terms in the second line of \eqref{trace} make no contribution since all the terms are 
proportional to $W\bar W$, and we have the nilpotency property $W_\alpha W_\beta W_\gamma=0$.
The square of the first term is easy to compute, and it is proportional to $W^2\bar{W}^2$.
Next we consider the cross product between the first and second term in  \eqref{trace} which leads to the relation: 
\be
W_\alpha \bar W_{\dot \alpha} \cdot D^\alpha W^2  \cdot \bar  D^{\dot \alpha}   \bar W^2
= W^2   (DW) \cdot \bar W^2(\bar D\bar W)=0
~.
\ee
Remarkably, this cross term vanishes since, as shown in Appendix \ref{appendix:EoMBI}, on-shell 
it is true that 
\be
W^2 \bar W^2 DW=0
~.
\label{on-shell_susy-condition}
\ee
A simple physical interpretation of this condition is that the manifest 
supersymmetry is preserved on-shell,  implying that the   auxiliary field $\sfD\propto D^\a W_\a|_{\q=0}$ 
has no vev,  and is at least linear in gaugino fields $\l_\a\propto W_{\a}|_{\q=0}$.
The vanishing of this cross term can be compared with the pure bosonic  case where the cross terms in $T^2$ 
vanish because of the tracelessness property of the free Maxwell stress tensor; see section \ref{bosonicBI}. 
 Finally, we  compute the square of the second term in \eqref{trace} which includes the following structure:
\be
 D^\alpha W^2  \cdot \bar  D^{\dot \alpha}  \bar W^2 \cdot  D_\alpha W^2  \cdot \bar  D_{\dot \alpha}  \bar W^2
= W^2 \bar W^2 D^2 W^2 \bar D^2 \bar W^2
~.
\ee
Here we have used $ (D_\alpha W_\beta)( D^\alpha W^\beta)  =  -\frac12 D^2 W^2  +W^\beta D^2 W_\beta$
to simplify the result.

In summary, on-shell the contributions to the supercurrent-squared operator $\cO_{T^2}$
defined in eq.~\eqref{superTsquare}
are given by    
\bsubeq\beqn
\mathcal J^2&=&
-\frac18  \Bigg\{ 
   \frac{16W^2\bar W^2}{\sqrt{1+2 \mathcal A+ \mathcal B^2}^2}+\frac{4W^2\bar W^2D^2 W^2 \bar D^2 \bar W^2 }{9 \sqrt{1+2 \mathcal A+ \mathcal B^2}^2  \Big( 1+\mathcal A  +\sqrt{ 1+2 \mathcal A+\mathcal  B^2    } \Big)^2 } 
   \Bigg\}
   ~,~~~~~~~~~
\\
\mathcal  X \bar \cX &=&\frac49     \frac{W^2\bar W^2 D^2 W^2 \bar D^2 \bar W^2 }
{  \sqrt{1+2 \mathcal A+ \mathcal B^2}  ^2\Big( 1+\mathcal A  +\sqrt{ 1+2 \mathcal A+\mathcal  B^2    } \Big)^2}
~.
\eeqn
\esubeq
Adding these results gives the supersymmetric $T^2$ primary operator $\cO_{T^2}$: 
\bsubeq
\beqn
  \mathcal O_{T^2}= -\frac{1}{2} \Big(    \mathcal J^2  +\frac54 \mathcal X \bar{\mathcal  X} \Big)
&=&  \frac{ W^2\bar W^2}{ \sqrt{1+2 \mathcal A+ \mathcal B^2}^2}
 \Bigg( 1-  \frac{ D^2 W^2 \bar D^2 \bar W^2 }
 { 4 \Big( 1+\mathcal A  +\sqrt{ 1+2 \mathcal A+\mathcal  B^2    } \Big)
^2 }   \Bigg)~,
~~~~~~~~~
\\&=&  \frac{W^2 \bar W^2}{   \sqrt{1+2 \mathcal A+ \mathcal B^2} ^2 } 
\Big( 1-  \frac{\mathcal A^2- \mathcal B^2 }{ (1+\mathcal A  +\sqrt{ 1+2 \mathcal A+\mathcal  B^2    })^2}  \Big)~,
\\&=&  \frac{2 W^2 \bar W^2}{ \sqrt{ 1+2 \mathcal A+\mathcal  B^2    }  \Big({1+\mathcal A  +\sqrt{ 1+2 \mathcal A+\mathcal  B^2 }  \Big)  }} 
~.
\label{OT2-susyBI}
\eeqn
\esubeq 
It is worth noting that the simplifications occurring in constructing $\cO_{T^2}$ from the supercurrents 
are very similar to the bosonic case of \eqref{OT2}.

Comparing with \eqref{BIflow}, we see that
eq.~\eqref{OT2-susyBI} proves that the supersymmetric BI action \eqref{susyBI1}
is an on-shell solution of the flow equation 
\bsubeq
\bea
\frac{\p  \mathcal L_\kappa}{\p \kappa^2} 
&=& 
 \int d^2 \theta d^2\bar \theta^2
 \,\frac{2 W^2 \bar W^2}{ \sqrt{ 1+2 \mathcal A+\mathcal  B^2    }  \Big({1+\mathcal A  +\sqrt{ 1+2 \mathcal A+\mathcal  B^2 }  
\Big)  }}~,
\\
&=& \int d^2 \theta d^2\bar \theta^2\, \mathcal O_{T^2} +\text{total derivatives} +{\rm EOM}
~.
\eea
\esubeq
It therefore describes a supercurrent-squared deformation
of the $\cN=1$ free Maxwell Lagrangian. 
This result establishes a relationship between non-linearly realized supersymmetry and 
supercurrent-squared flow equations in $D=4$.

Before closing  
 this section, we should make a few comments regarding the on-shell condition \eqref{on-shell_susy-condition}
 used in establishing the 
supercurrent-squared flow equation for the $D=4$ $\cN=1$ BI action.
First it is important to stress that the flow equation is not satisfied by the supersymmetric BI action off-shell. 
Second, we note that the specific combination of $ \mathcal J^2$ and $\mathcal X \bar{\mathcal  X}$ studied
is the unique choice for which \eqref{susyBI1} satisfies a supercurrent-squared flow equation, even if only on-shell.

Such a non-trivial condition satisfied by the on-shell supersymmetric BI action is intriguing and hints at the existence
of appropriate (super)field redefinitions which might
lead to a different supersymmetric extension of BI that satisfies the flow equation
off-shell. For example, it is know that the dependence of the off-shell extension on the  auxiliary field $\sfD$
can be modified by appropriate (super)field redefinitions, as well as redefinitions of the full superspace Lagrangian. 
We refer to \cite{Cecotti:1986gb, GonzalezRey:1998kh,Kuzenko:2011tj,Bagger:1997pi} for a list of relevant papers on this subject.
Under field redefinitions, the hidden supersymmetry will be modified but will remain a non-linearly realized symmetry of the 
theory. The existence of an off-shell solution of the supercurrent-squared flow is an interesting question for future research.


\section{\texorpdfstring{$D=4$}{D=4} Goldstino Action From Supercurrent-Squared Deformation}
\label{Goldstino-Flow}

In section \ref{BI-flows} we showed that the Bagger-Galperin action for the $D=4$ $\cN=1$ supersymmetric BI
theory satisfies a supercurrent-squared flow. It is known that the truncation of this model to fermions describes 
a Goldstino action for $D=4$ $\cN=1$ supersymmetry breaking; see, for example, \cite{Hatanaka:2003cr,Kuzenko:2005wh,Kuzenko:2011tj}. 
The $\cN=1$ non-linearly
realized supersymmetry arises as the non-linearly realized part of the $\cN=2\to\cN=1$ 
breaking of the supersymmetric BI.
We have shown in sections \ref{bosonicBI} and \ref{bosonicBI-2} 
that the bosonic truncation of the supersymmetric BI satisfies a $T^2$ flow equation.
The same should be true for the fermionic truncation.
More generally, one might argue that $D=4$ $\cN=1$ Goldstino models could satisfy a sort of flow equation that organizes their 
expansion in the supersymmetry breaking scale parameter. 

Note that in the $D=2$ case, the intuition is similar. If we consider the actions analyzed in section \ref{2D} 
that describe Goldstone models for partial $D=2$ $\cN=(4,4)\to\cN=(2,2)$ supersymmetry breaking,
one can immediately argue that their fermionic truncation describes Goldstino actions possessing non-linearly realized $D=2$
$\cN=(2,2)$ supersymmetry. These, by construction, are expected to satisfy a $T\Tb$-flow equation.
In fact,
such an argument is in agreement 
with the very nice recent analysis of \cite{Cribiori:2019xzp} 
where a $D=2$ Goldstino model possessing $\cN=(2,2)$ non-linearly realized supersymmetry
was shown to satisfy the supercurrent-squared flow equation \eqref{sc2_flow}.\footnote{We refer to 
\cite{Farakos:2016zam} for a discussion of various models possessing non-linearly realized $(2,2)$ supersymmetry.}
The model analyzed in \cite{Cribiori:2019xzp} 
is the analogue of the $D=4$ model  of \cite{Casalbuoni:1988xh,Komargodski:2009rz} and related on-shell to 
the Goldstino model of \cite{Rocek:1978nb}.\footnote{Note that the Goldstino models of
\cite{Casalbuoni:1988xh,Komargodski:2009rz,Rocek:1978nb} 
were shown in \cite{Kuzenko:2005wh,Kuzenko:2011tj} to be identical to the 
fermionic truncation of the supersymmetric BI action up to a field redefinition of the Goldstino.} 
This section is devoted to  showing that these $D=4$ $\cN=1$ Goldstino models 
satisfy a supercurrent-squared flow driven by the operator $\cO_{T^2}$ of the supersymmetric BI, in agreement with
the arguments given above.

 \subsection{\texorpdfstring{$D=4$}{D=4} Goldstino actions}  
   
The Volkov-Akulov (VA) action is the low energy description of supersymmetry breaking. There are several representations of the  
Goldstino action that are equivalent to the Volkov-Akulov form; see \cite{Kuzenko:2011tj,Cribiori:2016hdz} for  comprehensive discussions. 
Here we will focus on two models, but we start by reviewing a few general features of Goldstino actions.

The original VA action was obtained by requiring its invariance under the  the non-linear supersymmetry
 transformation~\cite{Volkov:1973ix}
 \be
 \delta_\xi \lambda^\alpha =\frac{1}{\kappa} \xi^\alpha -i \kappa(\lambda \sigma^m \bar \xi  -\xi \sigma^m \bar \lambda)\p_m \lambda^\alpha
~. \ee
 Explicitly, the original  Lagrangian was proven to be
 \be\label{VAaction}
 \mathcal L_{\rm VA}= -\frac{1}{2\kappa^2} \det A =
 -\frac{1}{2\kappa^2} -\frac{i}{2} 
 ( \lambda \sigma^m\p_m \bar \lambda-\p_m\lambda \sigma^m \bar \lambda  )
 + \text{interactions}
 ~,
 \ee  
 where 
 \be
 A_m{}^a= \delta_m{}^a -i \kappa^2 \p_m \lambda \sigma^a \bar \lambda +i \kappa^2 \lambda \sigma^a \p_m \bar \lambda
 ~.
 \ee
 
The alternative representation of the Goldstino action 
that interests us 
was originally introduced by Casalbuoni {\it et al.} in \cite{Casalbuoni:1988xh},
and later rediscovered and made fashionable
 by Komargodski and Seiberg \cite{Komargodski:2009rz}.
This model, which following recent literature we will call the KS model,
was constructed by imposing  nilpotent superfield constraints as a generalization of Ro\v{c}ek's 
seminal ideas for the Goldstino model described in \cite{Rocek:1978nb}. 
After integrating out an auxiliary field in the KS model, described in more detail in the next section, 
the  explicit form of the Lagrangian  is given by 
the following very simple combination of terms:
    \be\label{KSaction}
 \mathcal L_{\text{KS}}  = -f^2
-\frac{i}{2} 
 ( \psi \sigma^m\p_m \bar \psi-\p_m\psi \sigma^m \bar \psi  )
-  \frac{1}{4f^2} \p^\mu\bar \psi^2 \p_\mu \psi^2 -\frac{1}{16f^6} \psi^2 \bar\psi^2 \p^2 \psi^2 \p^2 \bar\psi^2
~.
   \ee
   The action is invariant under a quite involved non-linearly realized supersymmetry transformation whose
   explicit form can be found in \cite{Kuzenko:2010ef,Kuzenko:2005wh}.
 The Goldstino actions described by \eqref{VAaction} and \eqref{KSaction} 
 prove to be equivalent off-shell up to a field redefinition \cite{Kuzenko:2010ef,Kuzenko:2005wh}.

\subsection{\texorpdfstring{$D=4$}{D=4} KS Goldstino model as a supercurrent-squared flow}

 The goal in the rest of this section is to straightforwardly generalize   the analysis of \cite{Cribiori:2019xzp} to $D=4$ and to show how the  KS action satisfies a flow equation 
arising from a $T^2$ deformation of the free fermion action.

   \subsubsection{KS model}
   
Let us start by reviewing the Goldstino model of  \cite{Casalbuoni:1988xh,Komargodski:2009rz}.
Consider the following Lagrangian 
\be\label{KSlag}
\mathcal  L_{\rm KS}=\int d^4 \theta\, \bar \Phi \Phi
+ \int d^2 \theta  \Big( f  \Phi+ \frac12 \Lambda \Phi^2\Big) 
+ \int d^2 \bar\theta \Big( f \bar \Phi+   \frac12   \bar\Lambda \bar \Phi^2\Big)
~,
\ee
where $\Phi,\, \bar \Phi$ are $D=4$ $\cN=1$ chiral and anti-chiral superfields, satisfying the constraints $\bar D_{\dot \alpha  }\Phi=D_{\alpha} \bar \Phi=0$. The constant parameter $f$, which describes the supersymmetry breaking scale,
 is real. The superfields $\Lambda,\,\bar \Lambda$ are chiral and anti-chiral
  Lagrange multipliers whose ${\rm EOM}$ yield the nilpotent constraints
\be\label{nilpotent}
\Phi^2=\bar \Phi^2=0
~.
\ee
The equation of motion for $\Phi$ is
   \be\label{eomPhi}
   \frac14 \bar D^2 \bar \Phi=\Lambda \Phi+f
   ~, \qquad    \frac14  D^2  \Phi= \bar\Lambda \bar \Phi+f
   ~.
   \ee
As a consequence, we also have 
   \be\label{eomPhiDPhi}
   \Phi  \bar D^2 \bar \Phi =4f \Phi
   ~,\qquad      \bar \Phi  D^2   \Phi =4f \bar\Phi
   ~,
   \ee
 where the nilpotent properties of  \eqref{nilpotent} are used. 
 Note that the constraints \eqref{nilpotent} and \eqref{eomPhiDPhi} are the ones originally used by Ro\v{c}ek to define 
 his Goldstino model \cite{Rocek:1978nb}. These observations make manifest the on-shell equivalence of the KS model with Ro\v{c}ek's Goldstino model in a simple superspace setting. 
 The off-shell equivalence of all these Golstino models up to field redefinitions, including the VA action, was proven in 
 \cite{Kuzenko:2011tj}.

 The  Lagrange  multiplier  in  \eqref{KSlag} imposes the nilpotent constraint $\Phi^2=0$ on the chiral superfield $\Phi$. 
 This condition can be solved in terms of the spinor field $\psi$ and the auxiliary field $F$ of the chiral multiplet,  
 \cite{Casalbuoni:1988xh,Komargodski:2009rz}: 
    \be
   \Phi=\frac{\psi^2}{2F}+\sqrt{2} \theta \psi+\theta^2 F
   ~,
   \ee
   which is sensible assuming that $F\ne 0$.
Substituting back into \eqref{KSlag} gives a Lagrangian expressed  in terms of  $\psi $ and the auxiliary field $F$,
\be
\cL_{\rm KS}
=
-\frac{i}{2}\psi \s^\mu\pa_\mu\bar\psi
+\hf\bar{F}F
+\frac{1}{8}\frac{\bar\psi^2}{\bar{F}}\pa^2\Big(\frac{\psi^2}{F}\Big)
+fF
+c.c.~.
\label{KSaction2}
\ee
The auxiliary field can then be eliminated using its equation of motion, which can be solved in closed form
\be
F=-f\Bigg(
1
+\frac{\bar\psi^2}{4f^4}\pa^2\psi^2
-\frac{3}{16f^8}\psi^2\bar\psi^2\pa^2\psi^2\pa^2\bar\psi^2
\Bigg)
~,
\label{Fsolution}
\ee
together with the complex conjugate expression for $\bar{F}$.
Plugging \eqref{Fsolution} into \eqref{KSaction2}
gives the Goldstino action \eqref{KSaction}
 \cite{Casalbuoni:1988xh,Komargodski:2009rz}.

\subsubsection{$D=4$ Goldstino action as a supercurrent-squared flow}
   
One advantage of using the KS model compared to other Goldstino actions is the relatively simple form of the action, thanks to the Lagrange multiplier, which makes the computation of its supercurrent easier. 
The FZ multiplet  resulting from the action \eqref{KSlag} is 
\bsubeq
   \beqn
   \cJ_{\alpha\dot\alpha} &= & 
   2 D_\alpha\Phi \cdot \bar D_{\dot\alpha}   \bar \Phi -\frac23 [D_\alpha, \bar D_{\dot\alpha}] (\Phi \bar \Phi)
   =
   \frac23 D_\alpha \Phi  \cdot \bar D_{\dot\alpha} \bar \Phi 
-     \frac{2i}{3} \Big(    \Phi \p_{\alpha\dot\alpha} \bar \Phi -\bar \Phi \p_{\alpha\dot\alpha} \Phi \Big) 
~,~~~~~~
   \\
\cX   &= & 4\Big(f\Phi +\frac12\Lambda \Phi^2 \Big) 
 -\frac 13 \bar D^2 (\Phi \bar\Phi)
= \frac83 f\Phi +2\Lambda \Phi^2
~.
   \eeqn
 \esubeq
The composite operators $   \cJ^{\alpha\dot\alpha}   \cJ_{\alpha\dot\alpha}  $
and 
$\cX  \bar\cX$
are then
   \be
   \cJ^{\alpha\dot\alpha}   \cJ_{\alpha\dot\alpha}  = \frac{64}{9} f^2 \Phi\bar \Phi 
   + \text{total derivatives}
   +{\rm EOM}
   ~,
      \ee
and 
\be
\cX  \bar\cX  =\frac{64}{9}f^2 \Phi\bar\Phi
   +{\rm EOM}
~,
\ee   
where we used \eqref{nilpotent} and \eqref{eomPhi}.
The supercurrent-squared operator   \eqref{superTsquare}    then takes the form
   \be
     \mathcal O_{T^2}=  \frac{1}{16}  \cJ^{\alpha\dot\alpha}   \cJ_{\alpha\dot\alpha}- \frac58 \cX\bar \cX
     =-4f^2 \Phi\bar \Phi+ {\rm EOM}+  \text{total derivatives}
     ~.
   \ee
The descendant operator ${\sf O}_{T^2}$ of eq.~\eqref{superTsquare2} becomes
   \be
  {\sf O}_{T^2}  =\int d^2 \theta d^2 \bar \theta\;     \mathcal O_{T^2}
   =-4f^2\int d^2 \theta d^2 \bar \theta\;    \Phi\bar \Phi
   =2 f^3 \int d^2 \theta\, \Phi+2 f^3 \int d^2  \bar \theta\, \bar \Phi
   ~,
   \ee
  where we used \eqref{eomPhiDPhi} in the last equality.

  From \eqref{KSlag}, it is easy to see that the following relation holds:
  \be
  \frac{ \p\mathcal L_{\rm KS} }{\p f}=   \int d^2 \theta \,\Phi+  \int d^2  \bar \theta\, \bar \Phi
  ~.
 \ee
By identifying the coupling constants, 
   \be
\gamma=-\frac{1}{4   f^2 }
~,
   \ee
it follows immediately that the KS action,  
\be\label{SKS}
S_{\gamma}=\int d^4 x \, \mathcal L_{\rm KS}
~,
\ee
satisfies the flow equation
 \be\label{TTflow}
  \frac{ \p S_{\gamma} }{\p \gamma}=    \int d^4 x d^2 \theta\, \Phi+  \int d^4 x d^2   \bar \theta \,\bar \Phi 
  =\int d^4 x   d^2 \theta d^2\bar\theta  \;  \mathcal O_{T^2}
  = \int d^4 x \,  {\sf  O}_{T^2} 
  ~.
   \ee
   This proves that
   \eqref{KSlag} satisfies a supercurrent-squared flow
   (or $T^2$ flow) equation. 
  Because   on-shell the actions   \eqref{KSaction} and    \eqref{KSlag} are equivalent,
 and   the equation 
 $$\int  d^2 \theta d^2\bar\theta  \;  \mathcal O_{T^2}  =  {\sf  O}_{T^2} $$
 holds,
   eq.~\eqref{TTflow} proves that the $D=4$ $\cN=1$ Goldstino action arises from a supercurrent-squared deformation.\footnote{The careful reader may find that the flow can also be satisfied by other supercurrent-squared operators, $\cJ^2 - r \bar\cX\cX$, with arbitrary $r$ because of the linearity between $\cJ^2 $ and $\bar \cX \cX$. It is worth pointing out the same thing happens in $D=2$  \cite{Cribiori:2019xzp}. We stress that this is not the case for the supercurrent-squared flow satisfied by the $D=4$ supersymmetric Born-Infeld action.}

\section{Conclusions and Outlook}
\label{concludingthoughts}

In this work we have explored
the relationship between 
$T\Tb$ deformations and non-linear 
supersymmetry, extending the earlier analysis 
of 
\cite{Jiang:2019hux,Chang:2019kiu,Cribiori:2019xzp}. 
We first showed how two different $D=2$ $\cN=(2,2)$ 
$T\Tb$ deformations
of free supersymmetric scalar models, studied in~\cite{Chang:2019kiu}, classically possess a hidden non-linearly realized $\cN=(2,2)$ supersymmetry. The deformed theories are off-shell supersymmetric extensions of 
the gauge-fixed Nambu-Goto string in four dimensions. One way to potentially understand the appearance of non-linearly realized symmetries is by relating them to symmetries of the undeformed theories using the field-dependent change of variables discussed in~\cite{Conti:2018tca, Coleman:2019dvf}. 

These $D=2$ models turn out to be structurally 
very similar to the Bagger-Galperin action
describing a $D=4$ $\cN=1$ Born-Infeld theory, which possesses extra non-linearly realized $\cN=1$ supersymmetry \cite{Bagger:1996wp}. 
Inspired by this similarity and earlier work
on the  bosonic BI theory \cite{Conti:2018jho}, we proved that the $\cN=1$ BI action
satisfies a supercurrent-squared flow equation to all orders in the deformation parameter, extending the beautiful initial observation of \cite{cecotti:1987}.

Moreover, 
we concluded the paper by showing how the $D=4$ $\cN=1$ Goldstino action also
satisfies the same supercurrent-squared flow. This result extends the recent 
$D=2$ analysis of \cite{Cribiori:2019xzp}\ to four dimensions.
Our findings hint at an intriguing relation between current-squared deformations
and non-linear supersymmetry in various space-time dimensions that calls for a deeper explanation.

For the $D=2$ case where the $T\Tb$ operator 
is well-defined quantum mechanically,
it would be interesting to investigate other examples with various (super-)symmetry breaking patterns, and analyze the consistency conditions required by the existence of non-linear symmetries at the quantum level.\footnote{We are grateful to 
Guzm\'an Hern\'andez-Chifflet for stimulating comments on this subject.}

 For $D>2$, to the best of our knowledge, there is no complete argument showing 
that any of the proposed operators 
$O_{T^2}^{[r]}$ of
 eq.~\eqref{OT2r0}, including the holographic 
 operator of 
 \cite{Taylor:2018xcy, Hartman:2018tkw},
 possesses any particularly nice quantum properties.
 By looking at our $D=4$ $\cN=1$
 example, where the flow is controlled by
 the descendant operator 
 $ {\sf  O}_{T^2}$ of \eqref{superTsquare2},
 it seems clear that any
 supersymmetric completion of 
 $O_{T^2}^{[r]}$ will involve several other 
 current-squared operators. 
 An important question is to understand 
 whether such extensions have a hope of providing
 well-defined operators at the quantum level. This seems most
 promising in models with at least extended $\cN>1$, and more likely
 maximal, supersymmetry.

Putting aside the quantum properties 
of these deformations and flows, the connection 
between non-linear symmetries and $T\Tb$ flows might give
a novel way to organize 
interesting low-energy effective actions. The 
Born-Infeld and Goldstino actions that we have 
analyzed in this paper are universal low-energy structures in string theory, and in the latter case quantum field theory, 
precisely because of their non-linear symmetries, which
can be geometrically realized via brane physics.

The study of Volkov-Akulov-Dirac-Born-Infeld actions with extended supersymmetry in various space-time dimensions and their relationship to string theory has received a lot of attention in the past. We refer to the following (incomplete) list of references \cite{Cecotti:1986gb,Bagger:1996wp,Rocek:1997hi,Metsaev:1987by,GonzalezRey:1998kh,Metsaev:1987qp,Paban:1998ea,Paban:1998qy,Lin:2015ixa,Chen:2015hpa,Garousi:2017fbe,Heydeman:2017yww,Bergshoeff:1986jm,Tseytlin:1999dj,Bergshoeff:2013pia,Deser:1980ck,Ketov:2001dq,Kuzenko:2000tg,Kerstan:2002au,Ketov:1998ku,Ketov:1998sx,Ketov:2000zw,Kuzenko:2000uh,Bellucci:2000ft,Bellucci:2001hd,Bellucci:2000kc,Ivanov:2002ab,Berkovits:2002ag,Ivanov:2003uj}.
It would be remarkable if the $\alpha^\prime$
expansion of these models can be reorganized in a simple current-squared flow equation.
An efficient way to address the cases we have considered so far in $D=2$ and $D=4$ has been via superspace techniques. 
As a next step, one could try to analyze possible flow equations satisfied by the $D=4$ $\cN=2$ extensions of the DBI theory, which has been analyzed in superspace; see, for example,~\cite{Ketov:1998ku,Ketov:1998sx,Ketov:2000zw,Kuzenko:2000uh,Bellucci:2000ft,Bellucci:2001hd,Bellucci:2000kc}.

Another potentially tractable direction to be explored 
concerns the possible universality of the operator 
$ {\sf  O}_{T^2}$ of \eqref{superTsquare2} in the context of models with partial supersymmetry breaking. In the literature there are other known models for $D=4$ $\cN=2\to\cN=1$
supersymmetry breaking that share structural similarities with the Maxwell-Goldstone model of \cite{Bagger:1996wp}.
Well known are the Goldstone models based on the $D=4$ $\cN=1$ tensor multiplet \cite{Bagger:1997pi}, see also \cite{GonzalezRey:1998kh,Kuzenko:2011tj},
which have a dual description based on a chiral $\cN=1$ multiplet.
It is simple to show that at first order these actions satisfy a supercurrent-squared flow
analogous to the Bagger-Galperin action. Whether that result extends beyond leading order is
a natural question.

A final avenue for future investigation concerns the relationship between $T \Tb$ deformations and amplitudes. In two dimensions, $T \Tb$ simply modifies the $S$-matrix of the undeformed theory by a CDD factor \cite{Dubovsky:2013ira}, but one might wonder about the $S$-matrices of higher-dimensional theories deformed by generalizations of $T \Tb$. One hint is that theories with non-linearly realized symmetries exhibit enhanced soft behavior -- indeed, in the case of non-linearly realized \emph{super}symmetry, there is a proof that such symmetries generically lead to constraints on the soft behavior of the $S$-matrix \cite{Kallosh:2016qvo}, a fact which has been applied to the Volkov-Akulov action \cite{Kallosh:2016lwj}, which satisfies a $T \Tb$-like flow as we showed in section~\ref{Goldstino-Flow}. 

There are also examples involving purely bosonic theories. For instance, in four dimensions, the Dirac action is the unique Lorentz-invariant Lagrangian for a single scalar which is consistent with factorization, has one derivative per field, and exhibits soft degree $\sigma = 2$ for its scattering amplitudes \cite{Cheung:2014dqa}. Similarly, it has been shown that the Born-Infeld action for a vector can be fixed by demanding enhanced soft behavior in a particular multi-soft limit \cite{Cheung:2018oki}, which can be understood in the context of T-duality and dimensional reduction \cite{Elvang:2019twd}. Given the hints of a deeper relationship between supercurrent-squared deformations, non-linearly realized symmetries, and actions of Dirac or Born-Infeld type, it is natural to ask whether such deformations enhance the soft behavior of scattering amplitudes in a more general context.


\section*{Acknowledgements}

 We are  grateful 
to Chih-Kai Chang and Alessandro Sfondrini
for collaboration at early stages of 
this project. 
H.~J. especially would like to thank Ignatios Antoniadis for stimulating discussions and collaborations on related problems,  and  inspiration for  many of the ideas in the current project. 
C.~F., S.~S. and G.~T.-M.  would also like to thank the organizers and participants of the workshop on 
\emph{``$T \Tb$ and Other Solvable Deformations of Quantum Field Theories''} 
for providing a stimulating atmosphere, and the Simons Center for Geometry and Physics for hospitality and partial support 
during the initial stages of this work. C.~F., H.~J. and G.~T.-M. are also grateful for the support and 
vively atmosphere
during the workshop
\emph{``New frontiers of integrable deformations''}
in Villa Garbald, Castasegna 
during the final stage of preparation of this work.
C.~F. acknowledges support from the University of Chicago's divisional MS-PSD program. C.~F. and S.~S. are supported in 
part by NSF Grant No.~PHY 1720480. 
H.~J. is supported by Swiss National Science Foundation.
The work of G.~T.-M. was supported by the Albert Einstein Center for Fundamental Physics, University of Bern,
and by the Australian Research Council (ARC) Future Fellowship FT180100353.


 \appendix

\section{Deriving a Useful On-shell Identity}
\label{appendix:EoMBI}

This Appendix is devoted to deriving the on-shell relation \eqref{on-shell_susy-condition}.
We are going to prove this holds for an action of the form \eqref{GeneralBI}.
Let us start by considering the following Lagrangian
\bea
\cL&=&
\frac{1}{4}\int d^2\q\,W^2
+\frac{1}{4}\int d^2\bar{\q}\,\bar{W}^2
+\int d^2\q d^2\qb\,W^2\bar{W}^2\,\O\big[D^2W^2,\Db^2\bar{W}^2\big]
~.
\label{N=1_nl}
\eea
Remember that $W_\a$ and $\bar{W}_\ad$ satisfy the Bianchi identity
$D^\a W_\a=\Db_\ad \bar{W}^\ad$
whose solution is given in terms of a real but otherwise unconstrained scalar prepotential superfield $V$:
$W_\a=-1/4\,\Db^2D_\a V$
and
$\bar{W}_\ad=-1/4\,D^2\Db_\ad V$.
It is a straightforward calculation to derive the ${\rm EOM}$ 
by varying the action \eqref{N=1_nl} with respect to the prepotential $V$.
The ${\rm EOM}$ reads
\bea
    0&=&
-D^\a W_\a
+\frac{1}{2} D^\a\Db^2\Big(W_\a \bar{W}^2\O\Big)
+\frac{1}{2} \Db_\ad D^2\Big(W^2\bar{W}^\ad\O\Big)
\non\\
&&
+\frac{1}{2}D^\a\Big{[}  W_\a\Db^2D^2\Big(W^2\bar{W}^2\frac{\pa\O}{\pa(D^2W^2)}\Big)\Big{]}
+\frac{1}{2} \Db_\ad\Big{[}  \bar{W}^\ad \Big(D^2\Db^2W^2\bar{W}^2\frac{\pa\O}{\pa(\Db^2\bar{W}^2)}\Big)\Big{]}
~.~~~~~~
\label{EOM-N=1_nl}
\eea
Because of the constraint that $W_\a W_\b W_\g=0$ and its complex conjugate,
multiplying eq.~\eqref{EOM-N=1_nl} by $W^2\bar{W}^2$ and using the  
${\rm EOM}$ gives the following condition
\bea
&&
W^2\bar{W}^2(D^\a W_\a)\Big(1+f(\O)\Big)
=0
~,
\eea
where the functional $f(\O)$ is given by
\bea
f(\O)
&:=&
-\frac{1}{2}(\Db^2\bar{W}^2+D^2W^2)\O
\non\\
&&
-\frac{1}{2}\Bigg{[}
(D^2W^2)(\Db^2\bar{W}^2)\frac{\pa\O}{\pa(D^2W^2)}
+(D^2W^2)(\Db^2\bar{W}^2)\frac{\pa\O}{\pa(\Db^2\bar{W}^2)}
\Bigg{]}
~.
\eea
This implies
\bea
&&
W^2\bar{W}^2(D^\a W_\a)=0
~,
\eea
which is precisely condition \eqref{on-shell_susy-condition}.

\bibliographystyle{utphys}
\bibliography{master}

\providecommand{\href}[2]{#2}\begingroup\raggedright\begin{thebibliography}{10}

\bibitem{Zamolodchikov:2004ce}
A.~B. Zamolodchikov, ``{Expectation value of composite field T anti-T in
  two-dimensional quantum field theory},''
\href{http://www.arXiv.org/abs/hep-th/0401146}{{\tt hep-th/0401146}}.

\bibitem{Smirnov:2016lqw}
F.~A. Smirnov and A.~B. Zamolodchikov, ``{On space of integrable quantum field
  theories},'' {\em Nucl. Phys.} {\bf B915} (2017) 363--383,
\href{http://www.arXiv.org/abs/1608.05499}{{\tt 1608.05499}}.

\bibitem{Cavaglia:2016oda}
A.~Cavagli\`a, S.~Negro, I.~M. Sz\'ecs\'enyi, and R.~Tateo, ``{$T
  \bar{T}$-deformed 2D Quantum Field Theories},'' {\em JHEP} {\bf 10} (2016)
  112,
\href{http://www.arXiv.org/abs/1608.05534}{{\tt 1608.05534}}.

\bibitem{Dubovsky:2013ira}
S.~Dubovsky, V.~Gorbenko, and M.~Mirbabayi, ``{Natural Tuning: Towards A Proof
  of Concept},'' {\em JHEP} {\bf 09} (2013) 045,
\href{http://www.arXiv.org/abs/1305.6939}{{\tt 1305.6939}}.

\bibitem{Dubovsky:2017cnj}
S.~Dubovsky, V.~Gorbenko, and M.~Mirbabayi, ``{Asymptotic fragility, near
  AdS$_{2}$ holography and $ T\overline{T} $},'' {\em JHEP} {\bf 09} (2017)
  136,
\href{http://www.arXiv.org/abs/1706.06604}{{\tt 1706.06604}}.

\bibitem{Jiang:2019hxb}
Y.~Jiang, ``{Lectures on solvable irrelevant deformations of 2d quantum field
  theory},''
\href{http://www.arXiv.org/abs/1904.13376}{{\tt 1904.13376}}.

\bibitem{Bonelli:2018kik}
G.~Bonelli, N.~Doroud, and M.~Zhu, ``{$T \bar{T}$-deformations in closed
  form},'' {\em JHEP} {\bf 06} (2018) 149,
\href{http://www.arXiv.org/abs/1804.10967}{{\tt 1804.10967}}.

\bibitem{Dubovsky:2012wk}
S.~Dubovsky, R.~Flauger, and V.~Gorbenko, ``{Solving the Simplest Theory of
  Quantum Gravity},'' {\em JHEP} {\bf 09} (2012) 133,
\href{http://www.arXiv.org/abs/1205.6805}{{\tt 1205.6805}}.

\bibitem{Caselle:2013dra}
M.~Caselle, D.~Fioravanti, F.~Gliozzi, and R.~Tateo, ``{Quantisation of the
  effective string with TBA},'' {\em JHEP} {\bf 07} (2013) 071,
\href{http://www.arXiv.org/abs/1305.1278}{{\tt 1305.1278}}.

\bibitem{Chen:2018keo}
C.~Chen, P.~Conkey, S.~Dubovsky, and G.~Hern{\'a}ndez-Chifflet, ``{Undressing
  Confining Flux Tubes with $T\bar T$},'' {\em Phys. Rev.} {\bf D98} (2018),
  no.~11, 114024,
\href{http://www.arXiv.org/abs/1808.01339}{{\tt 1808.01339}}.

\bibitem{Dei:2018mfl}
A.~Dei and A.~Sfondrini, ``{Integrable spin chain for stringy
  Wess-Zumino-Witten models},'' {\em JHEP} {\bf 07} (2018) 109,
\href{http://www.arXiv.org/abs/1806.00422}{{\tt 1806.00422}}.

\bibitem{Baggio:2018gct}
M.~Baggio and A.~Sfondrini, ``{Strings on NS-NS Backgrounds as Integrable
  Deformations},'' {\em Phys. Rev.} {\bf D98} (2018), no.~2, 021902,
\href{http://www.arXiv.org/abs/1804.01998}{{\tt 1804.01998}}.

\bibitem{Frolov:2019nrr}
S.~Frolov, ``{TTbar deformation and the light-cone gauge},''
\href{http://www.arXiv.org/abs/1905.07946}{{\tt 1905.07946}}.

\bibitem{Sfondrini:2019smd}
A.~Sfondrini and S.~J. van Tongeren, ``{$T\bar{T}$ deformations as TsT
  transformations},''
\href{http://www.arXiv.org/abs/1908.09299}{{\tt 1908.09299}}.

\bibitem{Baggio:2018rpv}
M.~Baggio, A.~Sfondrini, G.~Tartaglino-Mazzucchelli, and H.~Walsh, ``{On $
  T\overline{T} $ deformations and supersymmetry},'' {\em JHEP} {\bf 06} (2019)
  063,
\href{http://www.arXiv.org/abs/1811.00533}{{\tt 1811.00533}}.

\bibitem{Chang:2018dge}
C.-K. Chang, C.~Ferko, and S.~Sethi, ``{Supersymmetry and $ T\overline{T} $
  deformations},'' {\em JHEP} {\bf 04} (2019) 131,
\href{http://www.arXiv.org/abs/1811.01895}{{\tt 1811.01895}}.

\bibitem{Jiang:2019hux}
H.~Jiang, A.~Sfondrini, and G.~Tartaglino-Mazzucchelli, ``{$T\bar{T}$
  deformations with $\mathcal{N}=(0,2)$ supersymmetry},'' {\em Phys. Rev.} {\bf
  D100} (2019), no.~4, 046017,
\href{http://www.arXiv.org/abs/1904.04760}{{\tt 1904.04760}}.

\bibitem{Chang:2019kiu}
C.-K. Chang, C.~Ferko, S.~Sethi, A.~Sfondrini, and G.~Tartaglino-Mazzucchelli,
  ``{$T\bar{T}$ Flows and (2,2) Supersymmetry},''
\href{http://www.arXiv.org/abs/1906.00467}{{\tt 1906.00467}}.

\bibitem{Ivanov:2000nk}
E.~Ivanov, S.~Krivonos, O.~Lechtenfeld, and B.~Zupnik, ``{Partial spontaneous
  breaking of two-dimensional supersymmetry},'' {\em Nucl. Phys.} {\bf B600}
  (2001) 235--271,
\href{http://www.arXiv.org/abs/hep-th/0012199}{{\tt hep-th/0012199}}.

\bibitem{Rocek:1997hi}
M.~Rocek and A.~A. Tseytlin, ``{Partial breaking of global D = 4 supersymmetry,
  constrained superfields, and three-brane actions},'' {\em Phys. Rev.} {\bf
  D59} (1999) 106001,
\href{http://www.arXiv.org/abs/hep-th/9811232}{{\tt hep-th/9811232}}.

\bibitem{Cribiori:2019xzp}
N.~Cribiori, F.~Farakos, and R.~von Unge, ``{The 2D Volkov-Akulov model as a $T
  \overline{T}$ deformation},''
\href{http://www.arXiv.org/abs/1907.08150}{{\tt 1907.08150}}.

\bibitem{Bagger:1996wp}
J.~Bagger and A.~Galperin, ``{A New Goldstone multiplet for partially broken
  supersymmetry},'' {\em Phys. Rev.} {\bf D55} (1997) 1091--1098,
\href{http://www.arXiv.org/abs/hep-th/9608177}{{\tt hep-th/9608177}}.

\bibitem{cecotti:1987}
S.~{Cecotti} and S.~{Ferrara}, ``{Supersymmetric born-infeld lagrangians},''
  {\em Physics Letters B} {\bf 187} (Mar., 1987) 335--339.

\bibitem{Conti:2018jho}
R.~Conti, L.~Iannella, S.~Negro, and R.~Tateo, ``{Generalised Born-Infeld
  models, Lax operators and the $ \mathrm{T}\overline{\mathrm{T}} $
  perturbation},'' {\em JHEP} {\bf 11} (2018) 007,
\href{http://www.arXiv.org/abs/1806.11515}{{\tt 1806.11515}}.

\bibitem{Coleman:2019dvf}
E.~A. Coleman, J.~Aguilera-Damia, D.~Z. Freedman, and R.~M. Soni, ``{$T
  \bar{T}$-Deformed Actions and (1,1) Supersymmetry},''
\href{http://www.arXiv.org/abs/1906.05439}{{\tt 1906.05439}}.

\bibitem{Ferrara:1974pz}
S.~Ferrara and B.~Zumino, ``{Transformation Properties of the Supercurrent},''
  {\em Nucl. Phys.} {\bf B87} (1975)
207.

\bibitem{Dumitrescu:2011iu}
T.~T. Dumitrescu and N.~Seiberg, ``{Supercurrents and Brane Currents in Diverse
  Dimensions},'' {\em JHEP} {\bf 07} (2011) 095,
\href{http://www.arXiv.org/abs/1106.0031}{{\tt 1106.0031}}.

\bibitem{Gates:1984nk}
S.~J. Gates, Jr., C.~M. Hull, and M.~Rocek, ``{Twisted Multiplets and New
  Supersymmetric Nonlinear Sigma Models},'' {\em Nucl. Phys.} {\bf B248} (1984)
157--186.

\bibitem{Buscher:1987uw}
T.~Buscher, U.~Lindstrom, and M.~Rocek, ``{New Supersymmetric $\sigma$ Models
  With {Wess-Zumino} Terms},'' {\em Phys. Lett.} {\bf B202} (1988)
94--98.

\bibitem{Grisaru:1997pg}
M.~T. Grisaru, M.~Massar, A.~Sevrin, and J.~Troost, ``{The Quantum geometry of
  N=(2,2) nonlinear sigma models},'' {\em Phys. Lett.} {\bf B412} (1997)
  53--58,
\href{http://www.arXiv.org/abs/hep-th/9706218}{{\tt hep-th/9706218}}.

\bibitem{Lindstrom:2005zr}
U.~Lindstrom, M.~Rocek, R.~von Unge, and M.~Zabzine, ``{Generalized Kahler
  manifolds and off-shell supersymmetry},'' {\em Commun. Math. Phys.} {\bf 269}
  (2007) 833--849,
\href{http://www.arXiv.org/abs/hep-th/0512164}{{\tt hep-th/0512164}}.

\bibitem{Ivanov:2004yv}
E.~Ivanov and A.~Sutulin, ``{Diverse N=(4,4) twisted multiplets in N=(2,2)
  superspace},'' {\em Theor. Math. Phys.} {\bf 145} (2005) 1425--1442,
  \href{http://www.arXiv.org/abs/hep-th/0409236}{{\tt hep-th/0409236}}.
[Teor. Mat. Fiz.145,66(2005)].

\bibitem{Kuzenko:2015rfx}
S.~M. Kuzenko and G.~Tartaglino-Mazzucchelli, ``{Nilpotent chiral superfield in
  N=2 supergravity and partial rigid supersymmetry breaking},'' {\em JHEP} {\bf
  03} (2016) 092,
\href{http://www.arXiv.org/abs/1512.01964}{{\tt 1512.01964}}.

\bibitem{Antoniadis:2017jsk}
I.~Antoniadis, J.-P. Derendinger, and C.~Markou, ``{Nonlinear $ \mathcal{N}=2 $
  global supersymmetry},'' {\em JHEP} {\bf 06} (2017) 052,
\href{http://www.arXiv.org/abs/1703.08806}{{\tt 1703.08806}}.

\bibitem{Antoniadis:2019gbd}
I.~Antoniadis, H.~Jiang, and O.~Lacombe, ``{$ \mathcal{N} $ = 2 supersymmetry
  deformations, electromagnetic duality and Dirac-Born-Infeld actions},'' {\em
  JHEP} {\bf 07} (2019) 147,
\href{http://www.arXiv.org/abs/1904.06339}{{\tt 1904.06339}}.

\bibitem{Gates:1983py}
S.~J. Gates, Jr., ``Superspace Formulation of New Nonlinear Sigma Models,''
  {\em Nucl. Phys.} {\bf B238} (1984)
349.

\bibitem{Gates:1995aj}
S.~J. Gates, Jr. and S.~V. Ketov, ``{2-D (4,4) hypermultiplets},'' {\em Phys.
  Lett.} {\bf B418} (1998) 111--118,
\href{http://www.arXiv.org/abs/hep-th/9504077}{{\tt hep-th/9504077}}.

\bibitem{Gates:1998fr}
S.~J. Gates and S.~V. Ketov, ``{2D(4,4) hypermultiplets. II: Field theory
  origins of dualities},'' {\em Phys. Lett.} {\bf B418} (1998)
119--124.

\bibitem{Ivanov:1984ht}
E.~A. Ivanov and S.~O. Krivonos, ``{N=4 SuperLiouville Equation (in
  Russian)},'' {\em J. Phys.} {\bf A17} (1984)
L671--L676.

\bibitem{Ivanov:1984fe}
E.~A. Ivanov and S.~O. Krivonos, ``{$N=4$ Superextension of the Liouville
  Equation With Quaternionic Structure},'' {\em Theor. Math. Phys.} {\bf 63}
  (1985) 477.
[Teor. Mat. Fiz.63,230(1985)].

\bibitem{Ivanov:1987mz}
E.~A. Ivanov, S.~O. Krivonos, and V.~M. Leviant, ``{A New Class of
  Superconformal $\sigma$ Models With the {Wess-Zumino} Action},'' {\em Nucl.
  Phys.} {\bf B304} (1988)
601--627.

\bibitem{Antoniadis:1995vb}
I.~Antoniadis, H.~Partouche, and T.~R. Taylor, ``{Spontaneous breaking of N=2
  global supersymmetry},'' {\em Phys. Lett.} {\bf B372} (1996) 83--87,
\href{http://www.arXiv.org/abs/hep-th/9512006}{{\tt hep-th/9512006}}.

\bibitem{Ivanov:1998jq}
E.~Ivanov and B.~Zupnik, ``{Modifying N=2 supersymmetry via partial
  breaking},'' in {\em {Theory of elementary particles. Proceedings, 31st
  International Symposium Ahrenshoop, Buckow, Germany, September 2-6, 1997}},
  pp.~64--69.
\newblock 1998.
\newblock
\href{http://www.arXiv.org/abs/hep-th/9801016}{{\tt hep-th/9801016}}.
\newblock

\bibitem{Taylor:2018xcy}
M.~Taylor, ``{TT deformations in general dimensions},''
\href{http://www.arXiv.org/abs/1805.10287}{{\tt 1805.10287}}.

\bibitem{Hartman:2018tkw}
T.~Hartman, J.~Kruthoff, E.~Shaghoulian, and A.~Tajdini, ``{Holography at
  finite cutoff with a $T^2$ deformation},'' {\em JHEP} {\bf 03} (2019) 004,
\href{http://www.arXiv.org/abs/1807.11401}{{\tt 1807.11401}}.

\bibitem{Cardy:2018sdv}
J.~Cardy, ``{The $ T\overline{T} $ deformation of quantum field theory as
  random geometry},'' {\em JHEP} {\bf 10} (2018) 186,
\href{http://www.arXiv.org/abs/1801.06895}{{\tt 1801.06895}}.

\bibitem{Komargodski:2010rb}
Z.~Komargodski and N.~Seiberg, ``{Comments on Supercurrent Multiplets,
  Supersymmetric Field Theories and Supergravity},'' {\em JHEP} {\bf 07} (2010)
  017,
\href{http://www.arXiv.org/abs/1002.2228}{{\tt 1002.2228}}.

\bibitem{Gates:1981yc}
S.~J. Gates, Jr., M.~T. Grisaru, and W.~Siegel, ``{Auxiliary Field
  Anomalies},'' {\em Nucl. Phys.} {\bf B203} (1982)
189--204.

\bibitem{Ambrosetti:2016ieg}
N.~Ambrosetti, D.~Arnold, J.-P. Derendinger, and J.~Hartong, ``{Gauge coupling
  field, currents, anomalies and $N=1$ super-Yang-Mills effective actions},''
  {\em Nucl. Phys.} {\bf B915} (2017) 285--334,
\href{http://www.arXiv.org/abs/1607.08646}{{\tt 1607.08646}}.

\bibitem{Gates:1983nr}
S.~J. Gates, M.~T. Grisaru, M.~Rocek, and W.~Siegel, ``Superspace, or one
  thousand and one lessons in supersymmetry,'' {\em Front. Phys.} {\bf 58}
  (1983) 1--548,
\href{http://www.arXiv.org/abs/hep-th/0108200}{{\tt hep-th/0108200}}.

\bibitem{Dienes:2009td}
K.~R. Dienes and B.~Thomas, ``{On the Inconsistency of Fayet-Iliopoulos Terms
  in Supergravity Theories},'' {\em Phys. Rev.} {\bf D81} (2010) 065023,
\href{http://www.arXiv.org/abs/0911.0677}{{\tt 0911.0677}}.

\bibitem{Kuzenko:2010am}
S.~M. Kuzenko, ``{Variant supercurrent multiplets},'' {\em JHEP} {\bf 04}
  (2010) 022,
\href{http://www.arXiv.org/abs/1002.4932}{{\tt 1002.4932}}.

\bibitem{Wess:1992cp}
J.~Wess and J.~Bagger, {\em {Supersymmetry and supergravity}}.
\newblock Princeton University Press, Princeton, NJ, USA,
1992.
\newblock

\bibitem{Rasheed:1997ns}
D.~A. Rasheed, ``{Nonlinear electrodynamics: Zeroth and first laws of black
  hole mechanics},''
\href{http://www.arXiv.org/abs/hep-th/9702087}{{\tt hep-th/9702087}}.

\bibitem{Deser:1980ck}
S.~Deser and R.~Puzalowski, ``{Supersymmetric Nonpolynomial Vector Multiplets
  and Causal Propagation},'' {\em J. Phys.} {\bf A13} (1980)
2501.

\bibitem{Kuzenko:2002vk}
S.~M. Kuzenko and S.~A. McCarthy, ``{Nonlinear selfduality and supergravity},''
  {\em JHEP} {\bf 02} (2003) 038,
\href{http://www.arXiv.org/abs/hep-th/0212039}{{\tt hep-th/0212039}}.

\bibitem{Cecotti:1986gb}
S.~Cecotti and S.~Ferrara, ``{Supersymmetric Born-Infeld Lagrangians},'' {\em
  Phys. Lett.} {\bf B187} (1987)
335--339.

\bibitem{GonzalezRey:1998kh}
F.~Gonzalez-Rey, I.~Y. Park, and M.~Rocek, ``{On dual 3-brane actions with
  partially broken N=2 supersymmetry},'' {\em Nucl. Phys.} {\bf B544} (1999)
  243--264,
\href{http://www.arXiv.org/abs/hep-th/9811130}{{\tt hep-th/9811130}}.

\bibitem{Kuzenko:2011tj}
S.~M. Kuzenko and S.~J. Tyler, ``{On the Goldstino actions and their
  symmetries},'' {\em JHEP} {\bf 05} (2011) 055,
\href{http://www.arXiv.org/abs/1102.3043}{{\tt 1102.3043}}.

\bibitem{Bagger:1997pi}
J.~Bagger and A.~Galperin, ``{The Tensor Goldstone multiplet for partially
  broken supersymmetry},'' {\em Phys. Lett.} {\bf B412} (1997) 296--300,
\href{http://www.arXiv.org/abs/hep-th/9707061}{{\tt hep-th/9707061}}.

\bibitem{Hatanaka:2003cr}
T.~Hatanaka and S.~V. Ketov, ``{On the universality of Goldstino action},''
  {\em Phys. Lett.} {\bf B580} (2004) 265--272,
\href{http://www.arXiv.org/abs/hep-th/0310152}{{\tt hep-th/0310152}}.

\bibitem{Kuzenko:2005wh}
S.~M. Kuzenko and S.~A. McCarthy, ``{On the component structure of N=1
  supersymmetric nonlinear electrodynamics},'' {\em JHEP} {\bf 05} (2005) 012,
\href{http://www.arXiv.org/abs/hep-th/0501172}{{\tt hep-th/0501172}}.

\bibitem{Farakos:2016zam}
F.~Farakos, P.~Koci, and R.~von Unge, ``{Superspace Higher Derivative Terms in
  Two Dimensions},'' {\em JHEP} {\bf 04} (2017) 002,
\href{http://www.arXiv.org/abs/1612.04361}{{\tt 1612.04361}}.

\bibitem{Casalbuoni:1988xh}
R.~Casalbuoni, S.~De~Curtis, D.~Dominici, F.~Feruglio, and R.~Gatto,
  ``{Nonlinear Realization of Supersymmetry Algebra From Supersymmetric
  Constraint},'' {\em Phys. Lett.} {\bf B220} (1989)
569--575.

\bibitem{Komargodski:2009rz}
Z.~Komargodski and N.~Seiberg, ``{From Linear SUSY to Constrained
  Superfields},'' {\em JHEP} {\bf 09} (2009) 066,
\href{http://www.arXiv.org/abs/0907.2441}{{\tt 0907.2441}}.

\bibitem{Rocek:1978nb}
M.~Rocek, ``{Linearizing the Volkov-Akulov Model},'' {\em Phys. Rev. Lett.}
  {\bf 41} (1978)
451--453.

\bibitem{Cribiori:2016hdz}
N.~Cribiori, G.~Dall'Agata, and F.~Farakos, ``{Interactions of N Goldstini in
  Superspace},'' {\em Phys. Rev.} {\bf D94} (2016), no.~6, 065019,
\href{http://www.arXiv.org/abs/1607.01277}{{\tt 1607.01277}}.

\bibitem{Volkov:1973ix}
D.~V. Volkov and V.~P. Akulov, ``{Is the Neutrino a Goldstone Particle?},''
  {\em Phys. Lett.} {\bf 46B} (1973)
109--110.

\bibitem{Kuzenko:2010ef}
S.~M. Kuzenko and S.~J. Tyler, ``{Relating the Komargodski-Seiberg and
  Akulov-Volkov actions: Exact nonlinear field redefinition},'' {\em Phys.
  Lett.} {\bf B698} (2011) 319--322,
\href{http://www.arXiv.org/abs/1009.3298}{{\tt 1009.3298}}.

\bibitem{Conti:2018tca}
R.~Conti, S.~Negro, and R.~Tateo, ``{The $ \mathrm{T}\overline{\mathrm{T}} $
  perturbation and its geometric interpretation},'' {\em JHEP} {\bf 02} (2019)
  085,
\href{http://www.arXiv.org/abs/1809.09593}{{\tt 1809.09593}}.

\bibitem{Metsaev:1987by}
R.~R. Metsaev and M.~Rakhmanov, ``{Fermionic Terms in the Open Superstring
  Effective Action},'' {\em Phys. Lett.} {\bf B193} (1987)
202--206.

\bibitem{Metsaev:1987qp}
R.~R. Metsaev, M.~Rakhmanov, and A.~A. Tseytlin, ``{The {Born-Infeld} Action as
  the Effective Action in the Open Superstring Theory},'' {\em Phys. Lett.}
  {\bf B193} (1987)
207--212.

\bibitem{Paban:1998ea}
S.~Paban, S.~Sethi, and M.~Stern, ``{Constraints from extended supersymmetry in
  quantum mechanics},'' {\em Nucl. Phys.} {\bf B534} (1998) 137--154,
\href{http://www.arXiv.org/abs/hep-th/9805018}{{\tt hep-th/9805018}}.

\bibitem{Paban:1998qy}
S.~Paban, S.~Sethi, and M.~Stern, ``{Supersymmetry and higher derivative terms
  in the effective action of Yang-Mills theories},'' {\em JHEP} {\bf 06} (1998)
  012,
\href{http://www.arXiv.org/abs/hep-th/9806028}{{\tt hep-th/9806028}}.

\bibitem{Lin:2015ixa}
Y.-H. Lin, S.-H. Shao, Y.~Wang, and X.~Yin, ``{Higher derivative couplings in
  theories with sixteen supersymmetries},'' {\em Phys. Rev.} {\bf D92} (2015),
  no.~12, 125017,
\href{http://www.arXiv.org/abs/1503.02077}{{\tt 1503.02077}}.

\bibitem{Chen:2015hpa}
W.-M. Chen, Y.-t. Huang, and C.~Wen, ``{Exact coefficients for higher
  dimensional operators with sixteen supersymmetries},'' {\em JHEP} {\bf 09}
  (2015) 098,
\href{http://www.arXiv.org/abs/1505.07093}{{\tt 1505.07093}}.

\bibitem{Garousi:2017fbe}
M.~R. Garousi, ``{Duality constraints on effective actions},'' {\em Phys.
  Rept.} {\bf 702} (2017) 1--30,
\href{http://www.arXiv.org/abs/1702.00191}{{\tt 1702.00191}}.

\bibitem{Heydeman:2017yww}
M.~Heydeman, J.~H. Schwarz, and C.~Wen, ``{M5-Brane and D-Brane Scattering
  Amplitudes},'' {\em JHEP} {\bf 12} (2017) 003,
\href{http://www.arXiv.org/abs/1710.02170}{{\tt 1710.02170}}.

\bibitem{Bergshoeff:1986jm}
E.~Bergshoeff, M.~Rakowski, and E.~Sezgin, ``{Higher Derivative Super
  Yang-Mills Theories},'' {\em Phys. Lett.} {\bf B185} (1987)
371--376.

\bibitem{Tseytlin:1999dj}
A.~A. Tseytlin, ``{Born-Infeld action, supersymmetry and string theory},''
\href{http://www.arXiv.org/abs/hep-th/9908105}{{\tt hep-th/9908105}}.

\bibitem{Bergshoeff:2013pia}
E.~Bergshoeff, F.~Coomans, R.~Kallosh, C.~S. Shahbazi, and A.~Van~Proeyen,
  ``{Dirac-Born-Infeld-Volkov-Akulov and Deformation of Supersymmetry},'' {\em
  JHEP} {\bf 08} (2013) 100,
\href{http://www.arXiv.org/abs/1303.5662}{{\tt 1303.5662}}.

\bibitem{Ketov:2001dq}
S.~V. Ketov, ``{Many faces of Born-Infeld theory},'' in {\em {7th International
  Wigner Symposium (Wigsym 7) College Park, Maryland, August 24-29, 2001}}.
\newblock 2001.
\newblock
\href{http://www.arXiv.org/abs/hep-th/0108189}{{\tt hep-th/0108189}}.
\newblock

\bibitem{Kuzenko:2000tg}
S.~M. Kuzenko and S.~Theisen, ``{Supersymmetric duality rotations},'' {\em
  JHEP} {\bf 03} (2000) 034,
\href{http://www.arXiv.org/abs/hep-th/0001068}{{\tt hep-th/0001068}}.

\bibitem{Kerstan:2002au}
S.~F. Kerstan, ``{Supersymmetric Born-Infeld from the D9-brane},'' {\em Class.
  Quant. Grav.} {\bf 19} (2002) 4525--4536,
\href{http://www.arXiv.org/abs/hep-th/0204225}{{\tt hep-th/0204225}}.

\bibitem{Ketov:1998ku}
S.~V. Ketov, ``{A Manifestly N=2 supersymmetric Born-Infeld action},'' {\em
  Mod. Phys. Lett.} {\bf A14} (1999) 501--510,
\href{http://www.arXiv.org/abs/hep-th/9809121}{{\tt hep-th/9809121}}.

\bibitem{Ketov:1998sx}
S.~V. Ketov, ``{Born-Infeld-Goldstone superfield actions for gauge fixed D-5
  branes and D-3 branes in 6-d},'' {\em Nucl. Phys.} {\bf B553} (1999)
  250--282,
\href{http://www.arXiv.org/abs/hep-th/9812051}{{\tt hep-th/9812051}}.

\bibitem{Ketov:2000zw}
S.~V. Ketov, ``{N=2 superBorn-Infeld theory revisited},'' {\em Class. Quant.
  Grav.} {\bf 17} (2000) L91,
\href{http://www.arXiv.org/abs/hep-th/0005126}{{\tt hep-th/0005126}}.

\bibitem{Kuzenko:2000uh}
S.~M. Kuzenko and S.~Theisen, ``{Nonlinear selfduality and supersymmetry},''
  {\em Fortsch. Phys.} {\bf 49} (2001) 273--309,
\href{http://www.arXiv.org/abs/hep-th/0007231}{{\tt hep-th/0007231}}.

\bibitem{Bellucci:2000ft}
S.~Bellucci, E.~Ivanov, and S.~Krivonos, ``{N=2 and N=4 supersymmetric
  Born-Infeld theories from nonlinear realizations},'' {\em Phys. Lett.} {\bf
  B502} (2001) 279--290,
\href{http://www.arXiv.org/abs/hep-th/0012236}{{\tt hep-th/0012236}}.

\bibitem{Bellucci:2001hd}
S.~Bellucci, E.~Ivanov, and S.~Krivonos, ``{Towards the complete N=2 superfield
  Born-Infeld action with partially broken N=4 supersymmetry},'' {\em Phys.
  Rev.} {\bf D64} (2001) 025014,
\href{http://www.arXiv.org/abs/hep-th/0101195}{{\tt hep-th/0101195}}.

\bibitem{Bellucci:2000kc}
S.~Bellucci, E.~Ivanov, and S.~Krivonos, ``{Superbranes and super-Born-Infeld
  theories from nonlinear realizations},'' {\em Nucl. Phys. Proc. Suppl.} {\bf
  102} (2001) 26--41, \href{http://www.arXiv.org/abs/hep-th/0103136}{{\tt
  hep-th/0103136}}.
[,26(2000)].

\bibitem{Ivanov:2002ab}
E.~A. Ivanov and B.~M. Zupnik, ``{New representation for Lagrangians of
  selfdual nonlinear electrodynamics},'' in {\em {Supersymmetries and Quantum
  Symmetries. Proceedings, 16th Max Born Symposium, SQS'01: Karpacz, Poland,
  September 21-25, 2001}}, pp.~235--250.
\newblock 2002.
\newblock
\href{http://www.arXiv.org/abs/hep-th/0202203}{{\tt hep-th/0202203}}.
\newblock

\bibitem{Berkovits:2002ag}
N.~Berkovits and V.~Pershin, ``{Supersymmetric Born-Infeld from the pure spinor
  formalism of the open superstring},'' {\em JHEP} {\bf 01} (2003) 023,
\href{http://www.arXiv.org/abs/hep-th/0205154}{{\tt hep-th/0205154}}.

\bibitem{Ivanov:2003uj}
E.~A. Ivanov and B.~M. Zupnik, ``{New approach to nonlinear electrodynamics:
  Dualities as symmetries of interaction},'' {\em Phys. Atom. Nucl.} {\bf 67}
  (2004) 2188--2199, \href{http://www.arXiv.org/abs/hep-th/0303192}{{\tt
  hep-th/0303192}}.
[Yad. Fiz.67,2212(2004)].

\bibitem{Kallosh:2016qvo}
R.~Kallosh, ``{Nonlinear (Super)Symmetries and Amplitudes},'' {\em JHEP} {\bf
  03} (2017) 038,
\href{http://www.arXiv.org/abs/1609.09123}{{\tt 1609.09123}}.

\bibitem{Kallosh:2016lwj}
R.~Kallosh, A.~Karlsson, and D.~Murli, ``{Origin of Soft Limits from Nonlinear
  Supersymmetry in Volkov-Akulov Theory},'' {\em JHEP} {\bf 03} (2017) 081,
\href{http://www.arXiv.org/abs/1609.09127}{{\tt 1609.09127}}.

\bibitem{Cheung:2014dqa}
C.~Cheung, K.~Kampf, J.~Novotny, and J.~Trnka, ``{Effective Field Theories from
  Soft Limits of Scattering Amplitudes},'' {\em Phys. Rev. Lett.} {\bf 114}
  (2015), no.~22, 221602,
\href{http://www.arXiv.org/abs/1412.4095}{{\tt 1412.4095}}.

\bibitem{Cheung:2018oki}
C.~Cheung, K.~Kampf, J.~Novotny, C.-H. Shen, J.~Trnka, and C.~Wen, ``{Vector
  Effective Field Theories from Soft Limits},'' {\em Phys. Rev. Lett.} {\bf
  120} (2018), no.~26, 261602,
\href{http://www.arXiv.org/abs/1801.01496}{{\tt 1801.01496}}.

\bibitem{Elvang:2019twd}
H.~Elvang, M.~Hadjiantonis, C.~R.~T. Jones, and S.~Paranjape,
  ``{All-Multiplicity One-Loop Amplitudes in Born-Infeld Electrodynamics from
  Generalized Unitarity},''
\href{http://www.arXiv.org/abs/1906.05321}{{\tt 1906.05321}}.

\end{thebibliography}\endgroup

\end{document}